\documentclass[fleqn,usenatbib]{mnras}

\usepackage{newtxtext,newtxmath}

\usepackage[T1]{fontenc}

\DeclareRobustCommand{\VAN}[3]{#2}
\let\VANthebibliography\thebibliography
\def\thebibliography{\DeclareRobustCommand{\VAN}[3]{##3}\VANthebibliography}

\usepackage{graphicx}	
\usepackage{amsmath}	
\usepackage{placeins}
\usepackage{xcolor}
\usepackage{lipsum}

\title[MeerKAT observations of Abell~1775 and~1795]{MeerKAT observations of Abell~1775 and Abell~1795: the discovery of a hadronic radio halo?}

\author[R. J. van Weeren et al.]{
R. J. van Weeren,$^{1}$\thanks{E-mail: rvweeren@strw.leidenuniv.nl (RJvW)}
           E.~Osinga,$^{2}$ G.~Brunetti,$^{3}$
           C.~J.~Riseley,$^{4,5}$
           A.~Botteon,$^{3}$
           R.~Timmerman,$^{6,7}$\newauthor
           A.~Bonafede,$^{8,3}$
           M.~Br\"uggen,$^{9}$
           R.~Cassano,$^{3}$
           V.~Cuciti,$^{8}$
           D.~Dallacasa,$^{8,3}$    
           F. de Gasperin,$^{3}$\newauthor
           J.~M.~G.~H.~J.~de~Jong,$^{1,10}$
           F.~Gastaldello,$^{11}$
           K.~Knowles,$^{12,13}$
           and X.~Zhang,$^{14}$
\\ \\
$^{1}$Leiden Observatory, Leiden University, PO Box 9513, 2300 RA Leiden, The Netherlands\\
$^{2}$Dunlap Institute for Astronomy \& Astrophysics, University of Toronto, 50 St. George Street, Toronto, ON M5S 3H4, Canada\\
$^{3}$INAF -- Istituto di Radioastronomia, via P. Gobetti 101, 40129 Bologna, Italy\\
$^{4}$Astronomisches Institut der Ruhr-Universit\"{a}t Bochum (AIRUB), Universit\"{a}tsstra{\ss}e 150, 44801 Bochum, Germany\\ 
$^{5}$Ruhr Astroparticle and Plasma Physics Center (RAPP Center), 44780 Bochum, Germany\\
$^{6}$Centre for Extragalactic Astronomy, Department of Physics, Durham University, Durham DH1 3LE, UK\\
$^{7}$Institute for Computational Cosmology, Department of Physics, Durham University, South Road, Durham DH1 3LE, UK\\
$^{8}$Dipartimento di Fisica e Astronomia, Universit\`a di Bologna, via P. Gobetti 93/2, I-40129 Bologna, Italy\\ 
$^{9}$University  of  Hamburg,  Hamburger  Sternwarte,  Gojenbergsweg 112, 21029 Hamburg, Germany\\ 
$^{10}$ASTRON, The Netherlands Institute for Radio Astronomy, Postbus 2, 7990 AA Dwingeloo, The Netherlands\\
$^{11}$INAF-IASF Milano, Via A. Corti 12, I-20133, Milano, Italy\\
$^{12}$Centre for Radio Astronomy Techniques and Technologies, Department of Physics and Electronics, Rhodes University, Makhanda 6140, South Africa\\
$^{13}$South African Radio Astronomy Observatory, Liesbeek House, River Park, Gloucester Road, Mowbray, Cape Town, 7700, South Africa\\
$^{14}$Max-Planck-Institut fur Extraterrestrische Physik, Giessenbachstrasse, 85748, Garching, Germany
}

\date{Accepted XXX. Received YYY; in original form ZZZ}

\pubyear{\the\year{}}

\begin{document}
\label{firstpage}
\pagerange{\pageref{firstpage}--\pageref{lastpage}}
\maketitle

\begin{abstract}

Giant radio haloes are diffuse synchrotron sources typically found in merging galaxy clusters, while smaller mini-haloes occur in cool-core clusters. Both trace cosmic-ray electrons in the intracluster medium, though recent observations suggest their distinction is not always clear. We present new 903--1655~MHz MeerKAT observations of Abell~1775 and Abell~1795, both hosting cool cores and cold fronts. Combined with reprocessed 120--168~MHz LOFAR Two-metre Sky Survey data, we perform imaging and spectral analyses of their radio emission.  In both clusters, we detect radio haloes with distinct inner and outer components. In Abell~1775, the halo appears diffuse at 1.3~GHz, while LOFAR images reveal steep-spectrum filaments. In Abell~1795, the inner component corresponds to a previously reported mini-halo candidate, but the full structure extends to $\sim$1~Mpc with a spectral index of $\alpha=-1.08\pm0.06$. The presence of such a large, flat-spectrum halo in a dynamically relaxed cluster makes Abell~1795 an outlier relative to typical merging systems. This suggests that some relaxed clusters may still retain sufficient turbulence to sustain particle re-acceleration, or that hadronic interactions producing secondary electrons play a significant role. Together with other recent discoveries in cool-core systems, our results indicate that some large radio haloes may have been overlooked in past studies due to limited dynamic range near bright central AGN. Finally, we detect steep-spectrum emission south of Abell~1795's central AGN, tracing a 45~kpc X-ray and optical filament that terminates in an X-ray cavity, likely linked to a past AGN outburst.

\end{abstract}




\section{Introduction}

Radio observations have revealed large-scale diffuse synchrotron-emitting sources in an increasing number of galaxy clusters \citep[see reviews by][]{2012A&ARv..20...54F,vanweeren+19}. This radio emission indicates the presence of magnetic fields and highly relativistic cosmic-ray electrons within the intracluster medium (ICM). Given the relatively short lifetimes of cosmic-ray electrons due to synchrotron and inverse Compton (IC) losses, these electrons must be produced in situ within the ICM \citep[e.g.,][]{1977ApJ...212....1J,2014IJMPD..2330007B}. Generally, cluster diffuse radio emission on scales of several hundred kiloparsec and larger is classified into two broad categories, radio relics (also called radio shocks) and radio haloes.

Here we will focus on radio haloes.  These are extended radio sources that, to first order, are co-located with the thermal X-ray emission from the hot ICM \citep[e.g.,][]{2001A&A...369..441G,2024A&A...686A...5B}. 
Giant radio haloes, with sizes of about 1--2~Mpc are typically found in unrelaxed merging galaxy clusters \citep[][]{2010ApJ...721L..82C,2023A&A...672A..43C,2023A&A...680A..30C}. The largest sample of radio haloes so far has been compiled from the LOFAR Two-metre Sky Survey \citep[LoTSS;][]{{2022A&A...659A...1S}} at a central frequency of 144~MHz \citep[][]{2022A&A...660A..78B}. It shows the presence of giant radio haloes in about $30\pm 11\%$ of clusters. Giant radio haloes have been explained by cosmic-ray electrons being re-accelerated by merging-induced turbulence \citep[][]{2001ApJ...557..560P,brunetti+01,brunetti+lazarian07}. This requires the presence of seed (``fossil'') cosmic rays in the ICM \citep[e.g.,][]{brunetti+01,2023A&A...669A..50V,2024Galax..12...19V}. In the turbulent re-acceleration model, there is a close link between the occurrence of radio haloes and cluster merger events \citep[][]{2010ApJ...721L..82C,
2023A&A...672A..43C,2021A&A...647A..51C}. The majority of radio haloes studied have spectral indices ($\alpha$, where $S_{\nu} \propto \nu^{\alpha}$ with $S_\nu$ the flux density at frequency $\nu$) of approximately $-1.1$ to $-1.3$. Moreover, radio haloes with ultra-steep spectra ($\alpha < -1.5$) are predicted to form in less energetic mergers \citep{2008Natur.455..944B}. In fact, the improved sensitivity of modern interferometers has allowed the discovery of a significant number of these haloes \citep[e.g.][]{2018MNRAS.473.3536W,2021A&A...654A.166D,2023A&A...669A...1R,2024A&A...689A.218P,2024ApJ...962...40S,2025arXiv250908062M}.

An alternative hadronic/secondary origin for radio haloes has also been proposed \citep[e.g.,][]{1980ApJ...239L..93D,1999APh....12..169B,2010ApJ...722..737K,2024MNRAS.527.1194K}. In this scenario, cosmic-ray electrons are produced as secondary particles from collisions between cosmic-ray protons and the thermal ICM. The observed properties of haloes and their connection with cluster dynamics disfavor the hadronic scenario \citep[reviews:][]{2014IJMPD..2330007B,vanweeren+19}. A remarkable prediction of the hadronic scenario is the gamma-rays generated by the decay of the neutral pions in the ICM. However, this emission has not been confirmed by Fermi-LAT observations \citep[e.g.,][]{2014ApJ...787...18A,2016ApJ...819..149A,Adam2021,2025MNRAS.539.2242L}; in particular, the faintness of the constraints on the gamma-ray signal has been used to argue against a purely hadronic origin of the haloes in the Coma and Abell~2256 clusters, once their spectral properties are taken into account \citep{2012MNRAS.426..956B,2017MNRAS.472.1506B,Adam2021,2024A&A...688A.175O}

Radio mini-haloes are found in the cores of relaxed (i.e., non-merging) galaxy clusters \citep[e.g.,][]{2017ApJ...841...71G}, with typical sizes of 200–400~kpc. The currently most favored scenarios are that they originate from the turbulent re-acceleration of cosmic ray electrons induced by sloshing motions in the cluster core \citep{2002A&A...386..456G,2011ApJ...743...16Z,2013ApJ...762...78Z} or from secondary electrons produced by hadronic collisions \citep[e.g.,][]{2004A&A...413...17P,2010ApJ...722..737K}. Recent observations indicate that mini-haloes and giant radio haloes can coexist in the same cluster \citep[e.g.,][]{2017A&A...603A.125V,2018MNRAS.478.2234S,2023A&A...678A.133B,2024A&A...692A..12V,2024A&A...686A..82B,2024A&A...686A..44R}. These cases may represent clusters where the merger does not (fully) disrupt the core, but can
inject enough turbulence at intermediate/large radii to re-accelerate cosmic rays and produce cluster wide diffuse emission. Being less energetic, such mergers could produce an ultra-steep-spectrum radio halo \citep[e.g.,][]{2018MNRAS.478.2234S,2024A&A...686A..82B}.

Recently, \cite{2024A&A...686A..82B} conducted a systematic study of cool-core clusters with LOFAR, finding that cluster-scale diffuse radio emission is not present in all cool-core clusters. However, its presence correlates with the existence of cold fronts. The detected emission is often asymmetric and follows the X-ray surface brightness distribution \citep[see also][]{2022MNRAS.512.4210R}. In some cases, the radio emission beyond the cluster core appeared confined to specific sectors, with the ``outer'' (giant) halo components visible only in certain regions. This may suggest that much of the giant halo component remained below the detection limit. Interestingly, \cite{2024ApJ...961..133G} showed that for two mini-halo hosting clusters, the emission beyond the cluster core was confined by more distant outer cold fronts, indicating that the entire diffuse emission appears to be related to large-scale sloshing.

The above result raises the question of whether our view of diffuse radio emission in cool-core clusters is complete. Several factors may hinder the detection of very extended radio emission in clusters with prominent cores. First, relaxed clusters have sharply peaked gas density profiles, meaning that radio emission is expected to fade rapidly with radius, given the known correlation between X-ray and radio surface brightness \citep[e.g.,][]{2001A&A...369..441G,2020A&A...640A..37I,2021A&A...654A..41R,2024A&A...683A.132L,2024A&A...686A...5B}. Additionally, cool-core clusters often host bright central AGN, making it challenging to detect faint diffuse emission -- Perseus being a prime example \citep{2024A&A...692A..12V}. Another possibility is that giant radio halo components in relaxed clusters have very steep spectra, making them harder to detect at high frequencies.

In this work, we present MeerKAT  \citep{2016mks..confE...1J} radio observations of two clusters with well-defined cores and confirmed cold fronts belonging to the larger Bo\"otes supercluster \citep{1997A&AS..123..119E,2014MNRAS.445.4073C}. Both clusters benefit from the availability of LOFAR observations. The first cluster, Abell~1775 ($z=0.0720$), is best known for the prominent tailed radio galaxy {B1339+266B} \citep[][]{1997ApJS..108...41O,2000NewA....5..335G,2007A&A...476...99G,2017A&A...608A..58T}. The cluster core also hosts several steep-spectrum filaments, along with a more diffuse halo component \citep[][]{2021A&A...649A..37B,2025arXiv250204913B}. Its dynamical state remains somewhat ambiguous, with \cite{2021A&A...649A..37B} suggesting that the cluster is undergoing a merger with a nonzero impact parameter, in which the core has not yet been fully disrupted. The second cluster, Abell~1795 ($z=0.0625$), is a strong cool-core cluster \citep[e.g.,][]{2002MNRAS.331..635E,2005MNRAS.361...17C,2015ApJ...799..174E} that exhibits evidence of sloshing motions \citep{2001ApJ...562L.153M}. A candidate mini-halo was reported by \cite{2014ApJ...781....9G}, but its presence could not be fully confirmed in GMRT, LOFAR, or MeerKAT observations \citep[][]{2018A&A...618A.152K,2022A&A...660A..78B,2023MNRAS.520.4410T}, with the central radio galaxy {4C\,+26.42} posing a significant challenge for a detection.

The outline of this paper is as follows.
The observations and data reduction are described in Sect.~\ref{sec:datareduction}. In  Sect.~\ref{sec:results}, the results are presented. We end the paper with a discussion and conclusions in Sects.~\ref{sec:discussion} and \ref{sec:conclusion}. We assume a $\Lambda$CDM cosmology with $H_{0} = 70$~km~s$^{-1}$~Mpc$^{-1}$, $\Omega_{m,0} = 0.3$, and $\Omega_{\Lambda,0} = 0.7$.

\section{Observations and data reduction}
\label{sec:datareduction}

\subsection{Observations}
For this work, we analyze MeerKAT observations of Abell\,1775 taken for project SCI-20220822-RV-01 (PI: van~Weeren) and archival observations for Abell~1795. Two of these observations were already presented in \cite{2023MNRAS.520.4410T}, but strong artifacts around the central radio source in the cluster limited the image quality, preventing the detection of diffuse emission. All observations were taken with the L-band receivers using the default 4k correlator mode, covering a digitised frequency range from 856 to 1712~MHz, with 4096 frequency channels (208.984~kHz per channel). The visibility integration time was set to 8~sec.  Table~\ref{tab:observations} presents an overview of the observations.

To complement the MeerKAT observations, we also include LOFAR 120–168 MHz data from the LOFAR Two-Metre Sky Survey \citep[LoTSS Data Release 2;][]{2022A&A...659A...1S}. These observations were previously presented in \cite{2021A&A...649A..37B} and \cite{2022A&A...660A..78B}. For Abell 1795, the LOFAR image quality was too low to confirm the presence of diffuse radio emission. In the case of Abell 1775, although a high-quality image was obtained, the cluster’s central region still exhibited some residual calibration artifacts. To improve the image quality for both targets, we reprocessed these observations with an improved calibration scheme and updated software
(see Sect.~\ref{sec:datareductionsub}).

\begin{table}
\addtolength{\tabcolsep}{-0.3em}
\begin{center}
{
\caption{MeerKAT observations}
\begin{tabular}{llllll} 
\hline
\hline
Cluster & Observing  & On-source & project-ID \\
&  dates & time [hr]\\
\hline
Abell 1775 &  {2022 Nov 22}  & 3.7 & SCI-20220822-RV-01\\
Abell 1795 &  {2019 Jun 16}  & 2.0 & SCI-20190418-KA-01\\
Abell 1795 &  {2019 Jun 17}  & 2.0 & SCI-20190418-KA-01\\ 
Abell 1795 &  {2021 Oct 16} & 1.4 & SCI-20210212-CR-01\\

\hline
\end{tabular}\label{tab:observations}
}
\end{center}
\end{table}

\subsection{Data reduction}
\label{sec:datareductionsub}

\subsubsection{MeerKAT}
For the MeerKAT data reduction, we started with the products from the SARAO Science Data Processor (SDP) continuum pipeline that were available in the archive. Here, we used the ``default calibrated'' option provided by the archive. This pipeline performs a standard calibration, including delay, bandpass, gain, and flux-scale calibration, as well as a conservative amount of flagging. When downloading, a factor of two averaging in frequency was performed. The target field data was subsequently split off with \texttt{CASA} \citep{2007ASPC..376..127M,2022PASP..134k4501C}. For the older Abell~1795 observations taken in 2019, no calibration was performed, and we bootstrapped the bandpass shape and flux scale from observations taken in 2021. Here, we used the full target field model derived from the 2021 observations. In addition, for the 2019 observations, two poorly performing antennas with low gain values were manually flagged.  

The rest of the data reduction was done with \texttt{facetselfcal} \citep[][]{2021A&A...651A.115V}, originally developed for self-calibrating LOFAR observations. \texttt{facetselfcal} is built upon the Default Pre-Processing Pipeline \citep[\texttt{DP3};][]{2018ascl.soft04003V} and \texttt{WSClean} for imaging \citep{2014MNRAS.444..606O,2017MNRAS.471..301O}, incorporating additional functionalities to further automate and improve self-calibration. Recent developments have also introduced features that enhance reproducibility in the updated versions of \texttt{facetselfcal} \citep{2025MNRAS.542.3253D}. It provides a flexible environment to perform self-calibration cycles with a range of user-optimised parameters, improving interoperability by allowing self-calibration of data from instruments other than LOFAR. A first demonstration of applying \texttt{facetselfcal} to MeerKAT data was presented by \cite{2024A&A...690A.222B}.

The visibility data were compressed with Dysco \citep{2016A&A...595A..99O}. Radio frequency interference (RFI) was flagged with \texttt{AOFlagger}  \citep{2010MNRAS.405..155O,2010ascl.soft10017O}, employing a specialised Stokes~Q, U, and V flagging strategy, which effectively removes RFI that is left in the SDP continuum pipeline data products. This strategy was updated from the one used in \cite{2024A&A...690A.222B}, which only did flagging based on Stokes~V. After flagging, noisy frequency channels near the edge of the band were removed due to the bandpass roll-off.  The output consists of data between 903 and 1655~MHz with 1800 equally spaced frequency channels.

The calibration process consisted of two main steps: direction-independent and direction-dependent (DD) calibration. For the direction-independent (DI) calibration, \texttt{DP3} `scalarphase' (single common phase solution for both polarizations ) self-calibration was performed on a full 3.3\degr\ field of view, using a solution interval of 1 minute and a frequency smoothness kernel of 100~MHz. At this stage, we did not apply amplitude self-calibration, as it generally did not lead to significant improvements in image quality. This is consistent with the findings of \cite{2020ApJ...888...61M}, indicating that most amplitude errors are due to  DD effects. For all imaging, we used \texttt{WSClean} with automatic clean masking. This was supplemented by a mask derived using \texttt{breizorro} \citep{2023ascl.soft05009R}, which avoids cleaning artifacts near bright sources. We employed multiscale cleaning \citep{2017MNRAS.471..301O} and multi-frequency synthesis with 12 frequency blocks. Following the initial full-field calibration, we performed an extraction step \citep{2021A&A...651A.115V}, where we subtracted sources outside a square region\footnote{2.5\degr{} and 1.7\degr{} for Abell~1775 and 1795, respectively} centred on the cluster. The size of this square was chosen to ensure that any sources causing remaining DD calibration artifacts affecting the cluster region were included. This subtraction was done by predicting the model visibilities of the sources and removing them from the visibility data.

In the final step, we performed DD calibration. The calibration directions were selected manually based on bright sources, with six directions for Abell~1775 and nine for Abell~1795. This faceted DD calibration was carried out in two stages: first, a `scalarphase' calibration with a 1~min solution interval, followed by a `scalarcomplexgain' (single common complexgain solution for both polarizations ) calibration with a 30~min solution interval, both using a frequency smoothness kernel of 100~MHz. 

For Abell~1795, the bright central source {4C\,+26.42}, associated with the brightest cluster galaxy (BCG), exhibited residual calibration artifacts. To address this, we refined the DD calibration solutions by reducing the `scalarphase' calibration solution interval to 16~sec for this direction. This approach leverages the new \texttt{DP3} functionality, which allows each direction to have a unique solution interval. Additionally, we introduced a third `scalarcomplexgain' solution for this direction with a 10~min calibration solution interval and a frequency smoothness kernel of 10~MHz. This strategy effectively mitigated the calibration issues caused by {4C\,+26.42}, which had previously complicated the analysis in \cite{2023MNRAS.520.4410T}. Furthermore, our results demonstrate that these challenges were not due to polarization leakage effects or the absence of secondary gain calibrations, as suspected by \cite{2023MNRAS.520.4410T}. Instead, they could be resolved using scalar (polarization-independent) DD corrections. 

Our resulting images are primary beam corrected using the model parameters described in \cite{2020ApJ...888...61M} and a central frequency of 1279~MHz. An overview of the image properties is given in Table~\ref{tab:imageproperties}.

\subsubsection{LOFAR}

The extracted LoTSS observations were calibrated using \texttt{facetselfcal}. These datasets, originally presented in \cite{2022A&A...660A..78B}, cover square regions of 33\arcmin{} and 52\arcmin{} for Abell~1775 and Abell~1795, respectively. For the calibration, we followed the general strategy of \citeauthor{2022A&A...660A..78B}, but employed an updated version of \texttt{facetselfcal}, which included several improvements--such as enhanced clean masking and the use of the \texttt{wgridder} algorithm in \texttt{WSClean} \citep{2021A&A...646A..58A,2022MNRAS.510.4110Y}. In addition, we increased the number of self-calibration cycles to 30, which led to noticeable further improvements in image quality compared to the 10 cycles normally carried out. This slow self-calibration convergence is partly due to LOFAR’s relatively poor outer $uv$-coverage outside the central LOFAR core. We also added a round of flagging with \texttt{AOflagger}, using the same strategy as for the MeerKAT data. For Abell~1795, this led to a significant reduction of artifacts around the bright central source {4C~+26.42}.

For both clusters, the ionospheric conditions were quite severe, with Abell~1795 being the most affected. These bad ionospheric conditions result in spoke-like patterns around bright sources. To mitigate these effects, we applied an additional DD calibration step to the output data from the direction-independent \texttt{facetselfcal} step described above. This involved calibrating in five directions for Abell~1795 and three for Abell~1775. Each facet typically contained at least one source with a peak flux of $\gtrsim$0.1~Jy. For Abell~1795, the final image shows a significant improvement over the data presented in \cite{2022A&A...660A..78B}. For Abell~1775, the improvements are more modest, as the original image artifacts were less severe \citep[see][]{2021A&A...649A..37B,2022A&A...660A..78B}. A comparison of these improvements with previously published images is presented in Appendix~\ref{sec:appendixpreviousLOFAR}.

\begin{figure*}
\centering
\includegraphics[width=0.49\textwidth]{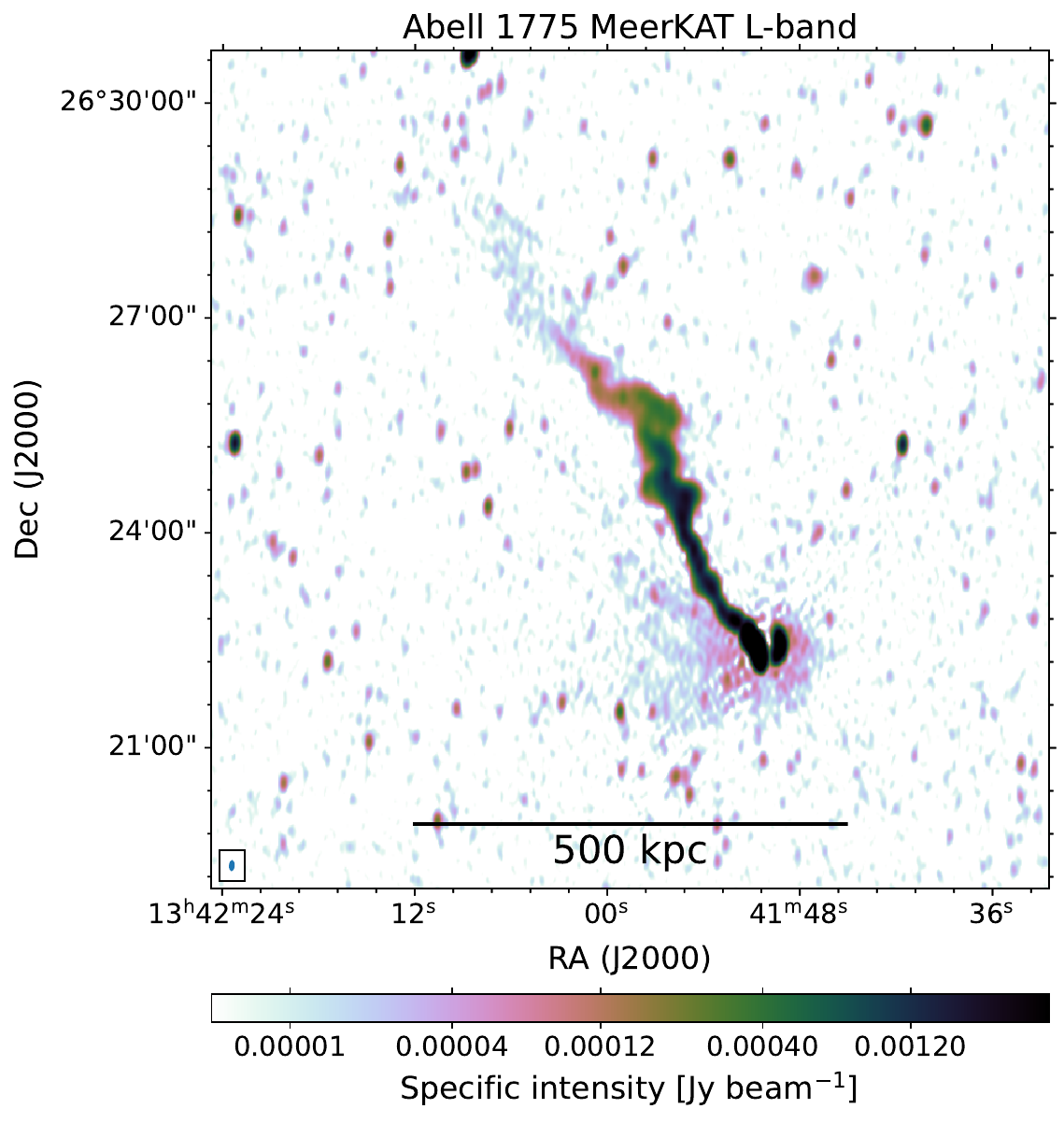}
\includegraphics[width=0.49\textwidth]{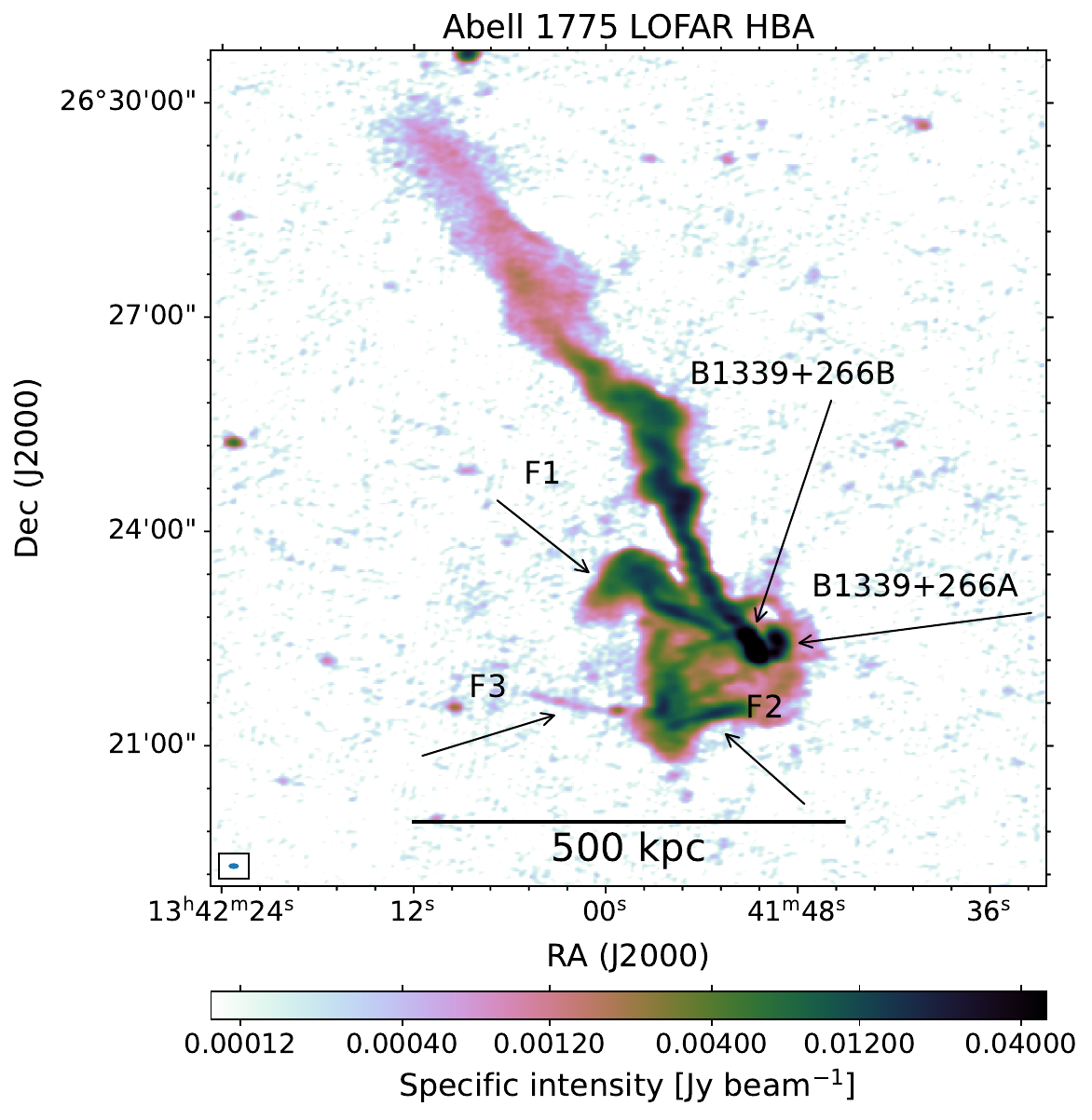}
\caption{MeerKAT L-band (left; at a resolution of $9.6\arcsec\times4.7\arcsec$) and LOFAR HBA (right; at a resolution of $9.1\arcsec\times5.0\arcsec$) images of Abell~1775 at central frequencies of 1279~MHz and 144~MHz, respectively. Images were made using Briggs weighting with a robust value of $-0.5$. Various features are labeled in the LOFAR image. The beam sizes are shown at the bottom left corners. The noise levels are reported in Table~\ref{tab:imageproperties}.}
\label{fig:radioA1775}
\end{figure*}

\begin{table}
\begin{center}
\caption{Image properties.}
 \setlength{\tabcolsep}{3pt}
\begin{tabular}{llllll} 
\hline
\hline
Image name      & Resolution,   & r.m.s noise & Figure\\
&               position angle$^\dagger$ & \\
&     \arcsec{} $\times$ \arcsec{}, \degr{} & $\mu$Jy\,beam$^{-1}$ \\
\hline
L-band A1775 & $9.6\times4.7$, 177 & 6.7 & \ref{fig:radioA1775} \\
L-band A1775 25~kpc &  $19.0\times18.5$, 6 & 9.4  & \ref{fig:MeerKATdiffuseA1775}\\
L-band A1775 50~kpc & $36.8\times36.4$, 140 & 20.0 & \ref{fig:MeerKATdiffuseA1775}, \ref{fig:lowres} \\
L-band A1775 100~kpc  &$73.0\times70.9$, 128 & 47.5 &\ref{fig:MeerKATdiffuseA1775},  \ref{fig:lowres}  \\
HBA A1775 &$9.1\times5.0$, 88 & 96.6 &  \ref{fig:radioA1775} \\
\hline
L-band A1795 &$10.4\times4.3$, 179 & 5.5 &  \ref{fig:radioA1795} \\
L-band A1795 25~kpc &$22.0\times21.0$, 167 & 9.7 & \ref{fig:radiodiffuseA1795} \\
L-band A1795 50~kpc &$42.4\times41.3$, 135 & 22.3 & \ref{fig:ChandraMeerKATA1795},  \ref{fig:lowres} \\
L-band A1795 100~kpc &$84.7\times80.3$, 128 & 60.7& \ref{fig:ChandraMeerKATA1795},  \ref{fig:lowres} \\
HBA A1795 &$9.4\times5.2$, 89 & 106 &  \ref{fig:radioA1795}  \\
HBA A1795 25~kpc& $28.9\times24.6$, 19 & 254 & \ref{fig:radiodiffuseA1795} \\
\hline
\end{tabular}\label{tab:imageproperties}
\end{center}
$^\dagger$ measured counter-clockwise from North toward East
\end{table}

\subsection{Compact source subtraction} 
\label{sec:subtraction}
To better highlight the diffuse radio emission, we produced low-resolution images with compact sources subtracted. First, we created a high-resolution image using only baselines that probe physical scales smaller than 100~kpc at the cluster's redshift. This was achieved by applying a minimum $uv$-cut during imaging with \texttt{WSClean}. The corresponding visibilities for this compact source model were then predicted and subtracted from the visibility data. Finally, we imaged the residual visibility data at a lower resolution, applying a Gaussian $uv$-taper corresponding to physical scales of 25, 50, and 100~kpc to improve the detectability of extended emission. 

\subsection{Spectral index maps}
We created a 144 -- 1279~MHz spectral index map by combining the MeerKAT L-band data with LOFAR HBA images. Both images were convolved to the same resolution and placed on the same pixel grid utilizing \texttt{CASA} before computing the spectral index maps. For both LOFAR and MeerKAT, the imaging was performed with a common inner $uv$-range cut of $80\lambda$. The uncertainty on the spectral index map was computed by adding the r.m.s. map noise and the absolute flux scale uncertainty in quadrature. For the absolute flux scale uncertainty, we adopt values of 10 and 5\% for LOFAR and MeerKAT, respectively \citep{2022A&A...659A...1S,2022MNRAS.512.4210R,2022A&A...659A.146D}. For the LOFAR images, we verified the flux scale alignment as described in \cite{2022A&A...657A...2R}, and found that for both clusters the required correction factors were smaller than the 10\% LOFAR flux-scale uncertainty. Therefore, no flux scale corrections were applied om the final images.

\section{Results}
\label{sec:results}

In the two subsections below, we will discuss the results of the two clusters, with a focus on the analysis of the diffuse radio emission associated with these systems.

\subsection{Abell~1775, PSZ2~G031.93+78.71}

\begin{figure*}
\centering
\includegraphics[width=0.47\textwidth]{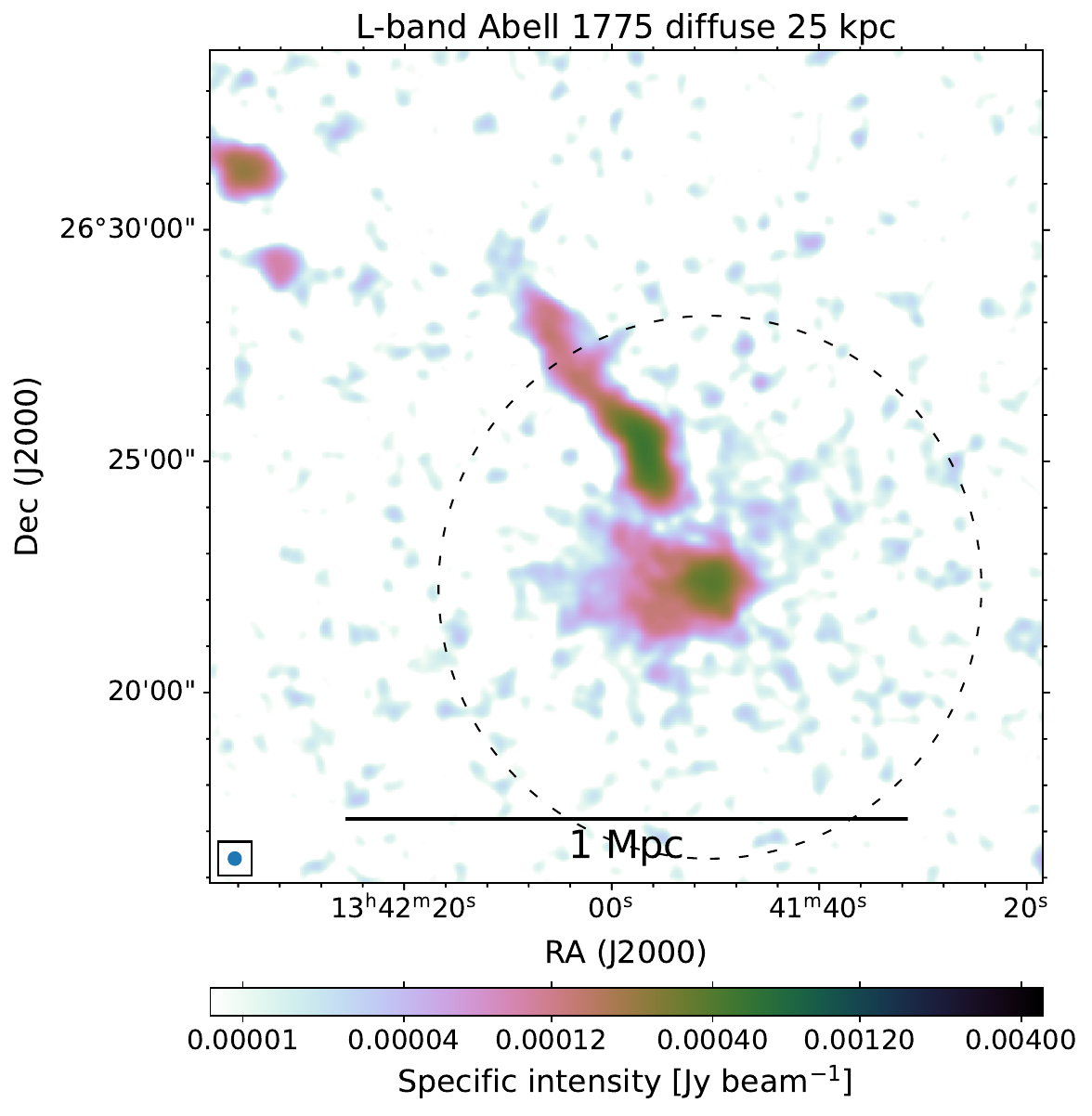}
\includegraphics[width=0.50\textwidth]{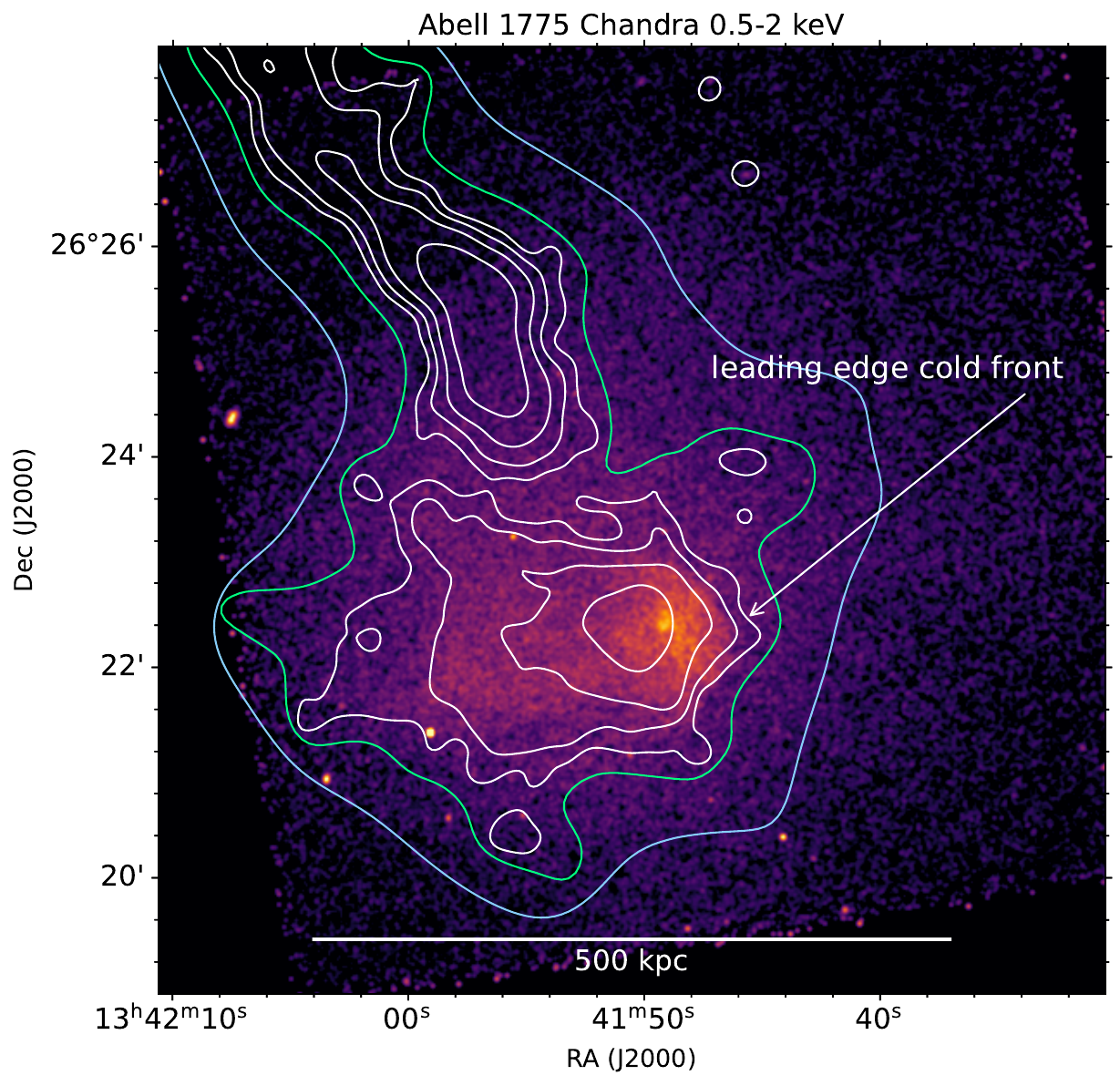}
\caption{Left panel: MeerKAT L-band image of Abell~1775 tapered to a resolution of 25~kpc at the cluster's redshift. The emission from compact sources was removed. The dashed circle indicates $0.5\times R_{500}$ \citep{2016A&A...594A..27P}. Right panel: Chandra 0.5--2.0~keV X-ray image of Abell~1775. MeerKAT L-band contours, with emission from compact sources subtracted, are overlaid. The two lowest contour levels come from images tapered at 100~kpc resolution (blue) and 50~kpc resolution (green), see also Fig.~\ref{fig:lowres}. These are drawn at a level of $5\times$ the r.m.s. map noise ($\sigma_{\rm{rms}}$). The white contours come from the 25~kpc resolution image shown in the left panel and are drawn at levels of $[1,2,4,\ldots] \times 5\sigma_{\rm{rms}}$.  The noise levels and beam sizes of the radio images are reported in Table~\ref{tab:imageproperties}.} 
\label{fig:MeerKATdiffuseA1775}
\end{figure*}

The MeerKAT and LOFAR images of Abell~1775 are shown in Fig.~\ref{fig:radioA1775}.
The radio emission in the cluster is dominated by the prominent tailed radio source {B1339+266B}, which is associated with the eastern giant elliptical galaxy of the cluster. The radio tail of {B1339+266B} extends toward the northeast. Just west of B1339+266B's nucleus lies the more compact, bright radio source {B1339+266A}, which is associated with a second giant elliptical cluster galaxy.

The MeerKAT image also reveals diffuse emission surrounding the cluster's core, enveloping the nuclei of {B1339+266B} and {B1339+266A}. This emission has a relatively smooth but asymmetric morphology, fading more slowly in the eastern direction. 
In the lower-resolution compact-source-subtracted image smoothed to 25~kpc displayed in Fig.~\ref{fig:MeerKATdiffuseA1775}, the asymmetric brightness distribution becomes even more apparent. An overlay with the Chandra image taken from \cite{2021A&A...649A..37B} reveals a striking correspondence between the diffuse radio and X-ray emission. We therefore classify the diffuse emission as a radio halo. Notably, the radio surface brightness drops at the location of the western ``leading edge'' cold front of the cluster's core identified by \cite{2021A&A...649A..37B}. On the eastern side, the radio emission follows the overall X-ray morphology, which appears to trace material stripped from the cluster core. In the 50 and 100~kpc resolution images (Fig.~\ref{fig:lowres}), more extended radio emission is observed, extending beyond the ``leading edge'' cold front. This large-scale emission is also somewhat elongated in the east-west direction, mirroring the broader X-ray structure (Fig.~\ref{fig:MeerKATdiffuseA1775} right panel). The total extent of the diffuse radio emission, excluding the contribution from the tailed source {B1339+266B}, is approximately 500~kpc. This larger-scale diffuse emission can be identified as the faint radio halo component previously reported by \cite{2021A&A...649A..37B}.

\begin{figure*}
\centering
\includegraphics[width=0.49\textwidth]{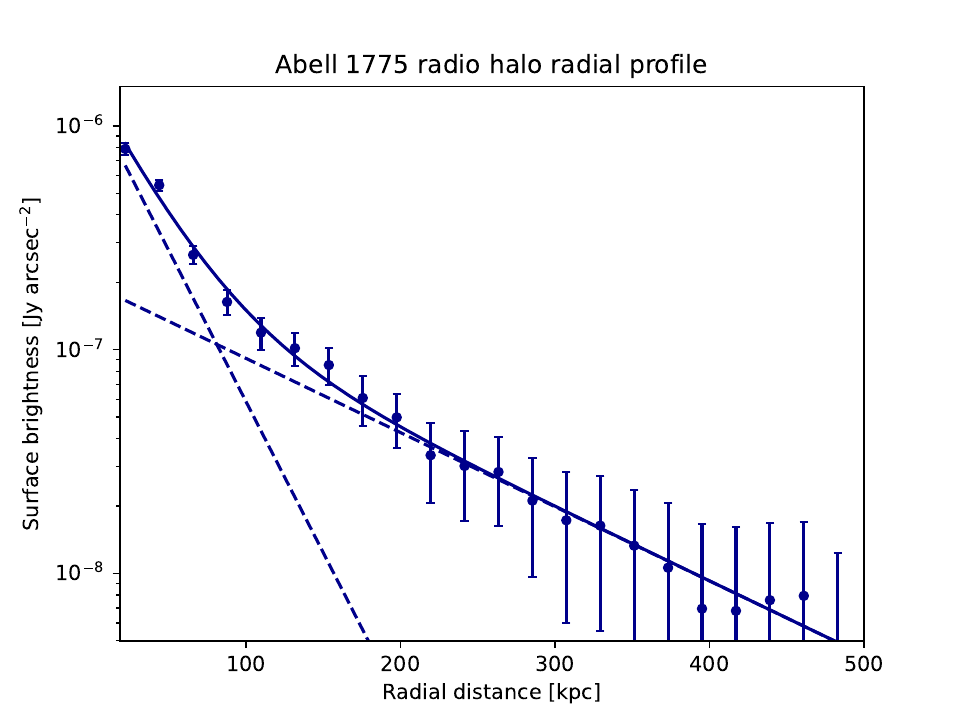}
\includegraphics[width=0.49\textwidth]{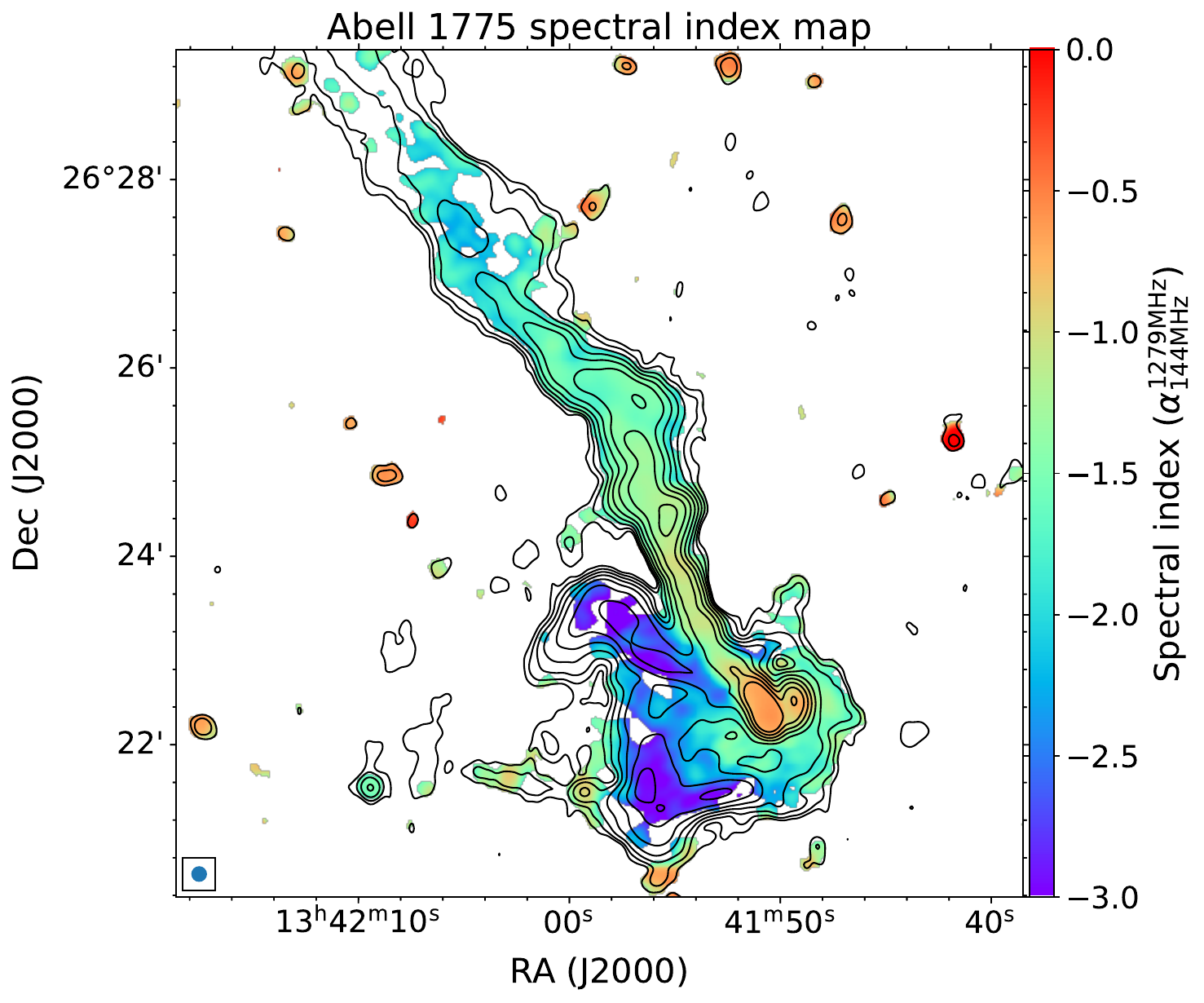}
\caption{Left panel: Radial radio surface brightness profile for Abell~1775 extracted from the MeerKAT 25~kpc resolution image. The solid blue line shows the best fitting double exponential model (Eq.~\ref{eq:profile}). The dashed lines show the two individual exponential components. Right panel: Spectral index map at 10\arcsec~resolution between 144 and 1279 MHz for Abell~1775. Contours are from the 144 MHz LOFAR image and are drawn at levels of $[1,2,4,\ldots] \times 3\sigma_{\rm{rms}}$, with $\sigma_{\rm{rms}}=119$~$\mu$Jy~beam$^{-1}$. The corresponding spectral index uncertainty map is shown in Fig.~\ref{fig:spixerror}.}
\label{fig:radioprofileA1775}
\end{figure*}

We extracted an azimuthally averaged radial radio surface brightness profile, centred on B1339+266B, to investigate the brightness distribution of the radio halo. For this, we used the 25~kpc resolution MeerKAT image with the emission from compact sources subtracted. In addition, we masked regions (indicated in Fig.~\ref{fig:lowres}) where the subtraction failed to remove the emission from unrelated extended sources, the most relevant one being the tail of {B1339+266B}, including the ``inner tail'' part \citep[see][]{2021A&A...649A..37B}. The resulting radial profile is shown in Fig.~\ref{fig:radioprofileA1775} (left panel). The profile displays two components, a steep inner component and a shallow, fainter outer component. Based on this profile, we fit a double exponential model to the data points \citep{2009A&A...499..679M,2021A&C....3500464B,2024A&A...692A..12V} of the form 
\begin{equation}
I_{\nu}\left(r\right) = I_{\nu,\rm{0,inner}} e^{-r/r_{\rm{e,inner}}} +  I_{\nu,\rm{0,outer}} e^{-r/r_{\rm{e,outer}}} \mbox{ .}
\label{eq:profile}
\end{equation} 
The model fits the observed data well, with the best fitting model parameters being reported in Table~\ref{tab:profile}. We do not find evidence for a shallower outer component at large radii, which can sometimes arise from the incomplete subtraction of compact sources \citep{2025arXiv250505415R}. We tested different masking strategies, but these did not lead to any significant changes in the derived profile.

From the fit, we obtain an integrated flux density of $15.8\pm2.5$~mJy for the entire two-component radio halo, integrating out to $r=\infty$ at 1279~MHz. Taking the commonly adopted 80\% of this value (corresponding to integrating to $r=3r_e$ in case of a single exponential model), we obtain $P_{\rm{1.4\mbox{ }GHz}} = \left(1.45\pm 0.23\right)\times 10^{23}$~W~Hz$^{-1}$, scaling with a typical radio halo spectral index of $\alpha=-1.3$ \citep[e.g.,][]{vanweeren+19}.  The characteristic radius of the outer component, $r_{\rm{e,outer}} = (131 \pm 42)$~kpc, falls within the range observed for giant radio haloes \citep[e.g.,][]{2022A&A...660A..78B}, although it lies toward the smaller end of the distribution.

\begin{table*}
\begin{center}
\caption{Radio halo profile modeling. The integrated flux densities ($S_{\rm{inner}}$, $S_{\rm{outer}}$, and $S_{\rm{total}}$) are calculated by integrating to $r=\infty$. The radio powers ($P_{\rm{inner}}$, $P_{\rm{outer}}$, $P_{\rm{total}}$) are calculated based on taking 80\% of these integrated flux density values,  corresponding with integrating to $r=3r_e$. The uncertainties on the flux densities and radio powers reported in the table include the absolute flux scale uncertainty, which was added in quadrature to the uncertainties from the profile fitting. The profile fitting accounted for the statistical uncertainties due to image noise.}

\begin{tabular}{llllll} 
\hline
\hline
& Abell~1775 -- 1279~MHz  & Abell~1795 -- 1279~MHz  & Abell~1795 -- 144~MHz  \\
\hline
$I_{\rm{0,inner}}$ [$\mu$Jy\,arcsec$^{-2}$] &  $1.32\pm0.16$ & $5.8\pm0.3$  & $75\pm8$\\
$I_{\rm{0,outer}}$ [$\mu$Jy\,arcsec$^{-2}$] &  $0.20\pm0.09$ & $0.60\pm0.07$& $3.9\pm1.5$\\
$r_{\rm{e,inner}}$ [kpc] &   $32.2\pm 4.3$  &  $38.6\pm0.8$ & $38.5\pm0.9$\\
$r_{\rm{e,outer}}$ [kpc] &   $131\pm42$     &  $135\pm7$    & $157\pm35$\\
$S_{\rm{inner}}$ [mJy] &     $4.5 \pm 1.1$  & $38.1\pm2.5$ & $490\pm62$ \\
$S_{\rm{outer}}$ [mJy] &     $11.2\pm 2.0$  & $48.6\pm3.1$ & $426\pm65$\\
$S_{\rm{total}}$ [mJy] &     $15.8 \pm 2.5$ & $86.7\pm4.7$ & $916\pm109$\\
$P_{\rm{inner}}$ [W~Hz$^{-1}$] & $\left(4.2 \pm 1.0 \right) \times 10^{22}$ & $\left(2.6 \pm 0.2 \right) \times 10^{23}$ & $\left(3.5 \pm 0.4 \right) \times 10^{24}$\\
$P_{\rm{outer}}$ [W~Hz$^{-1}$] & $\left(1.03\pm 0.19\right) \times 10^{23}$ & $\left(3.3 \pm 0.2 \right) \times 10^{23}$ & $\left(3.0 \pm 0.5 \right) \times 10^{24}$\\
$P_{\rm{total}}$ [W~Hz$^{-1}$] & $\left(1.45\pm 0.23\right) \times 10^{23}$ & $\left(5.8 \pm 0.3 \right) \times 10^{23}$ & $\left(6.5 \pm 0.8 \right) \times 10^{24}$\\
\hline
\end{tabular}\label{tab:profile}
\end{center}
\end{table*}

The LOFAR 144~MHz image of the cluster is shown in Fig.~\ref{fig:radioA1775} (right panel). It reveals a dramatically different picture in the cluster core compared to MeerKAT, with the emission dominated by a complex web of filaments. These ultra-steep spectrum filaments were first detected by \citep{2021A&A...649A..37B} and studied in more detail by \cite{2025arXiv250204913B}. We point out the presence of an additional narrow thin filament extending east \citep[also visible in the image presented][]{2021A&A...649A..37B}, with a length of ${\sim}100\arcsec$, corresponding to ${\sim}140$~kpc. We label this feature F3.

In Fig.~\ref{fig:radioprofileA1775} (right panel), we present the cluster's 144--1279 MHz spectral index map at 10\arcsec~resolution. The eastern parts of F1 and F2 exhibit an ultra-steep spectrum with $\alpha =-3.0 \pm 0.1$. 
Since the diffuse radio halo component dominates at L-band, the actual spectral index of F1 and F2 in these regions is likely even steeper. In the easternmost part of the filament complex, where no spectral index could be determined due to the absence of detectable emission in the MeerKAT image, we derive a $1\sigma$ upper limit on the spectral index of about $-1.5$.
In the cluster core, surrounding the nuclei of B1339+266A+B, the spectral index flattens to {$\alpha=-1.5\pm0.1$}. The spectral index of the diffuse halo emission seen in the MeerKAT image is difficult to determine because the steep-spectrum filaments, which dominate at low frequencies, contaminate the measurement.

\subsection{Abell~1795, PSZ2 G033.81+77.18}

\begin{figure*}
\centering
\includegraphics[width=0.49\textwidth]{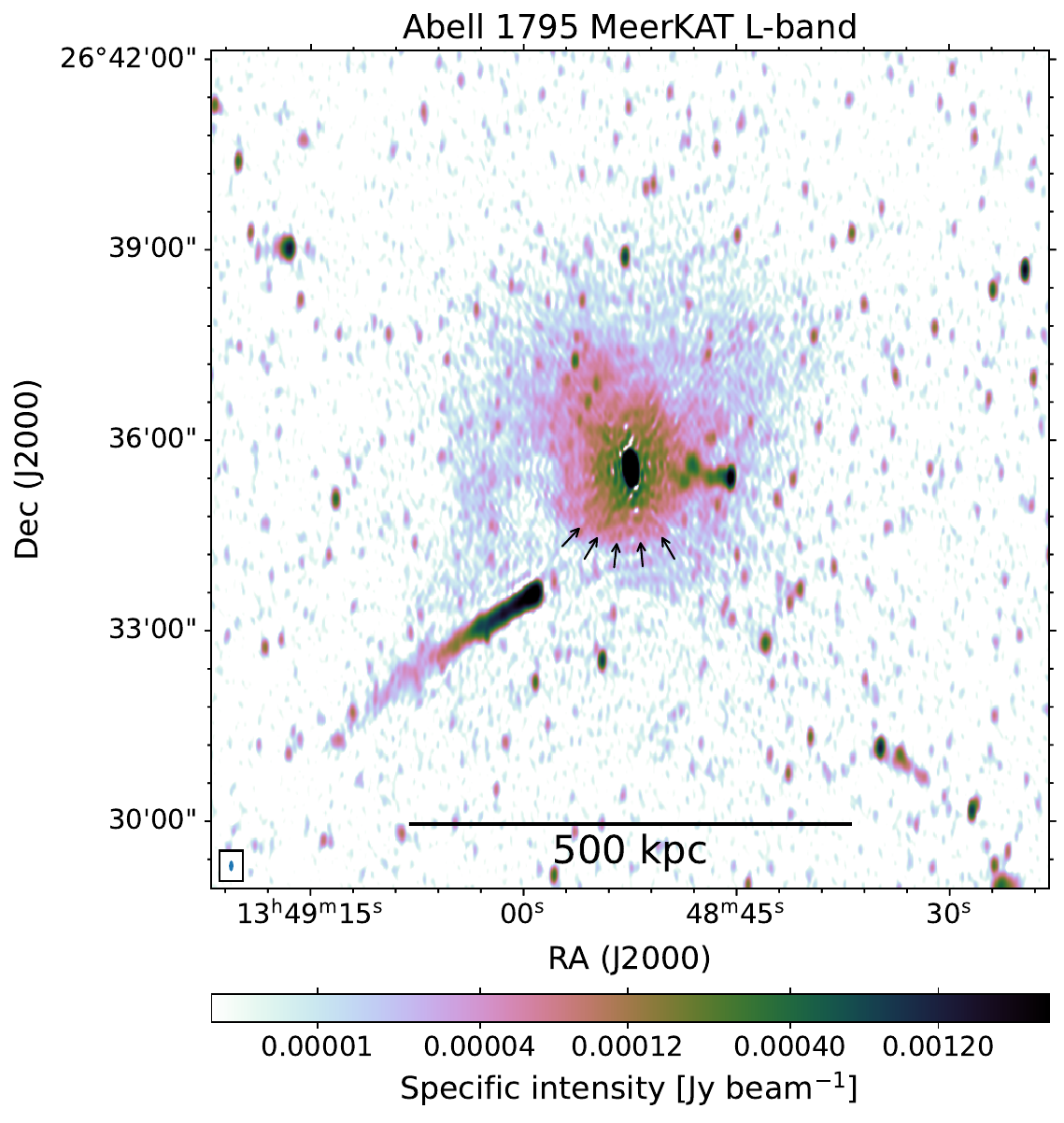}
\includegraphics[width=0.49\textwidth]{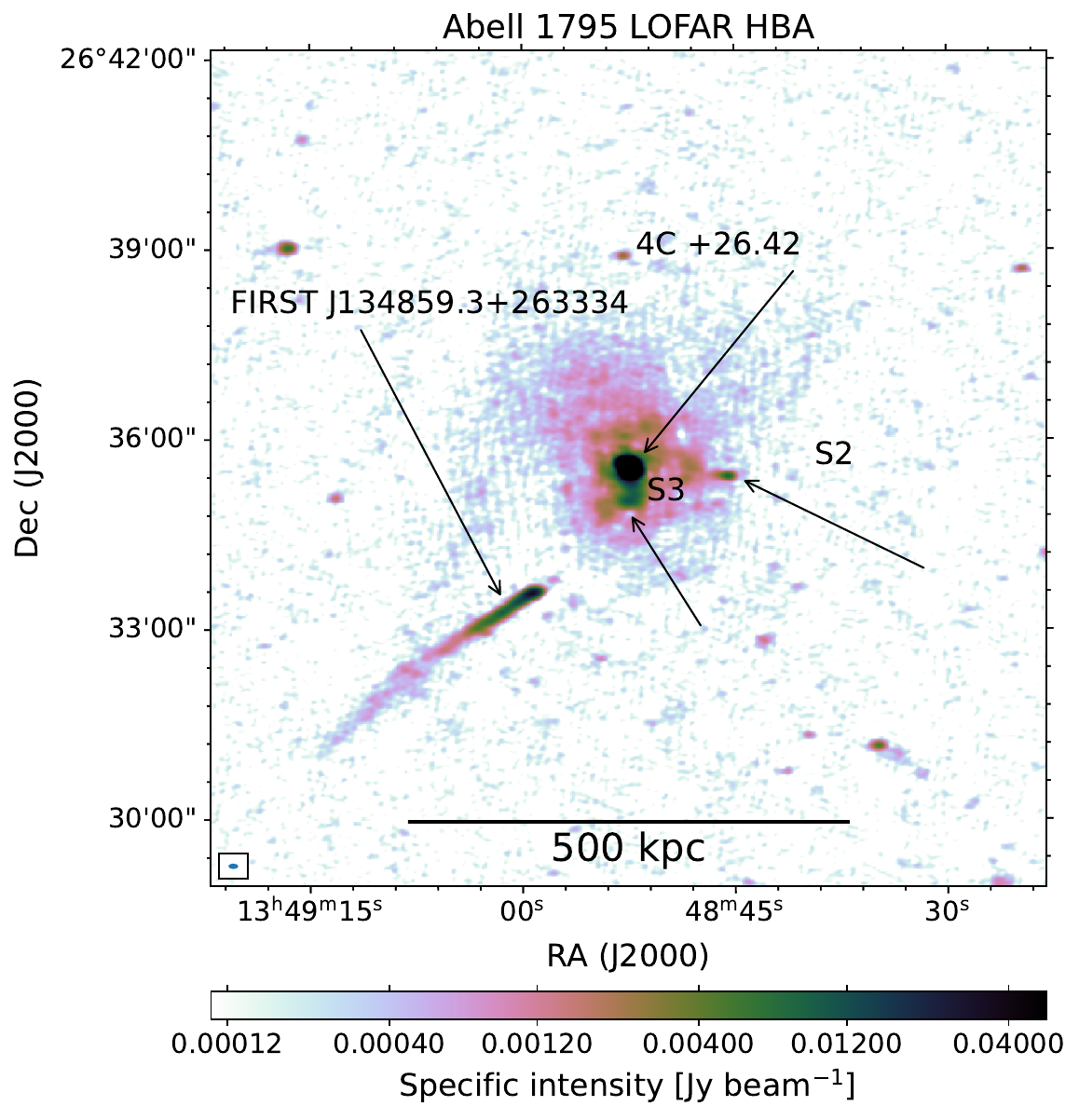}
\caption{MeerKAT L-band (left; at a resolution of $10.4\arcsec\times4.3\arcsec$) and LOFAR HBA (right; at a resolution of $9.4\arcsec\times5.2\arcsec$) images of Abell~1795 at central frequencies of 1279~MHz and 144~MHz, respectively. Images were made using Briggs weighting with a robust value of $-0.5$.  The arrows on the MeerKAT image indicate the location of the cold front \citep{2001ApJ...562L.153M}. Various radio features are labeled on the LOFAR image. The beam sizes are shown at the bottom left corners.  The noise levels are reported in Table~\ref{tab:imageproperties}.}
\label{fig:radioA1795}
\end{figure*}

The MeerKAT and LOFAR images of Abell~1795 are shown in Fig.~\ref{fig:radioA1795}. In contrast to Abell~1775, both images reveal the same general features. The compact powerful radio source  {4C\,+26.42} is associated with the dominant BCG \citep{1984ApJ...276...79V,1993AJ....105..778G,2018A&A...618A.152K}. The images also show the tailed radio galaxy {FIRST\,J134859.3+263334} to the southeast of the cluster core, with the tail pointing away from the cluster towards the southeast \citep{1993ApJS...87..135O,1997ApJS..108...41O,2017A&A...608A..58T}. To the west of the cluster's core, source S2 is located, where we follow the naming of \cite{2014ApJ...781....9G}. Extended tail-like emission is detected from S2 towards the east in the MeerKAT image. Given that S2 is associated with a distant background galaxy, as mentioned in \citeauthor{2014ApJ...781....9G}, it seems likely that this tail is not associated with Abell~1795 itself.

\begin{figure}
\centering
\includegraphics[width=0.49\textwidth]{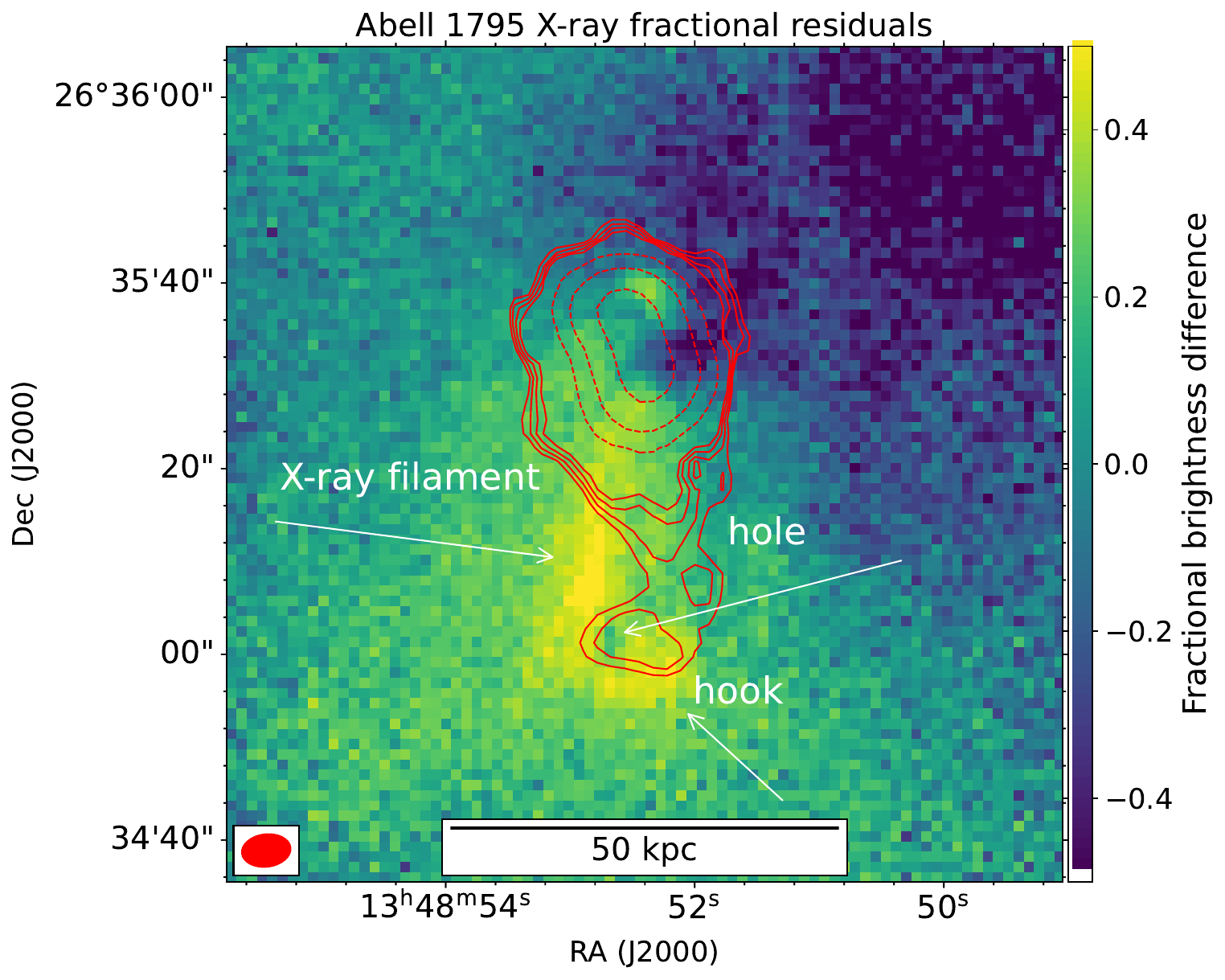}
\caption{\textit{Chandra} 0.5–-2.0~keV fractional residual X-ray image. The image was created by subtracting a radially averaged profile from the original X-ray image and dividing the result by the original. LOFAR radio contours at 144~MHz are overlaid in red. These contours are from a high-resolution image with a Briggs robust parameter of $-1.25$, yielding a beam size of $5.3\arcsec \times 3.5\arcsec$ with a position angle of 97\degr (shown in the bottom left corner). Solid contours are at levels of $[1, 2, 4, 8] \times 25\sigma_{\rm rms}$; dashed contours are at $[0.2, 1, 5] \times 10^{3}\sigma_{\rm rms}$, with $\sigma_{\rm rms} = 139$~$\mu$Jy~beam$^{-1}$.}
\label{fig:A1795ghost}
\end{figure}

A notable feature detected in the LOFAR image is located just south of {4C\,+26.42}, labeled S3 in the right panel of Fig.~\ref{fig:radioA1795}. This feature, which lies along the extension of the known southern {4C\,+26.42} lobe, has not been reported in previous studies, including the early LOFAR analysis by \cite{2020MNRAS.496.2613B}, and is not discernible in the MeerKAT image. We measure a spectral index of $-1.63 \pm 0.05$ for S3 between 1279 and 144 MHz. However, the actual spectral index is likely steeper, as the MeerKAT emission in this region appears to be entirely dominated by surrounding diffuse emission, with no clear contribution from S3.
Figure~\ref{fig:A1795ghost} displays a fractional residual X-ray image of the cluster centre, showing a known X-ray filament or tail that extends to the south \citep{2001MNRAS.321L..33F}. This filament traces cooler gas and displays a hook-like structure at its southern end \citep{2005MNRAS.361...17C,2018A&A...618A.152K}. A small X-ray depression or ``hole'' is located between the end of the filament and the hook. This feature was previously identified and discussed by \citet{2005MNRAS.361...17C, 2014MNRAS.445.3444W, 2018A&A...618A.152K}. The LOFAR image shows that S3 traces radio emission that extends southwards from {4C\,+26.42}, skirting the western edge of the X-ray filament and leading into the small-scale X-ray hole. We discuss the implications of this discovery for the origin of the X-ray filament and the associated depression in Sect.~\ref{sec:ghost}.

Surrounding {4C\,+26.42}, we find diffuse radio emission.  
Hints of this emission were seen by \cite{2014ApJ...781....9G,2023MNRAS.520.4410T} but could not be confirmed as a mini-halo. The diffuse emission shows a remarkable spiral-like pattern, with a trail of emission extending to the tailed galaxy {FIRST\,J134859.3+263334}.  
On the southern side of the cluster's core, the radio surface brightness emission drops sharply. This drop coincides with the location of the cold front identified by \cite{2001ApJ...562L.153M}. However, the halo emission is not confined by this cold front, with lower surface brightness emission found south of the cold front. The low-resolution MeerKAT and LOFAR images with emission from compact sources subtracted (see Fig.~\ref{fig:radiodiffuseA1795}) also show a slight elongation of the halo emission towards the northwest, marked as ``NW elongation''.  The NW elongation does not coincide with the cavity reported by \cite{2014MNRAS.445.3444W}, which is located at a smaller cluster-centric distance compared to where the radio elongation becomes apparent. 

In the LOFAR image, we detect a faint ``patch'' of emission located 9.5\arcmin~(680~kpc) northeast of the cluster centre, labeled ``NE patch'' in Fig.~\ref{fig:radiodiffuseA1795} (right panel). No obvious counterpart to this source is identified. This ``NE patch'' has an extent of 1.8\arcmin{} (corresponding to ${\sim}130$~kpc at the redshift of Abell~1795) and is not detected in the MeerKAT image. In the LOFAR image, smoothed to 25~kpc resolution, we measure an integrated flux density of $S_{144}=10.7\pm1.4$~mJy. The non-detection in the MeerKAT images implies a steep spectral index, with a $1\sigma$ upper limit of $\alpha <-2.1$, calculated by summing the flux in this region from the ~25~kpc resolution MeerKAT image and using that as the upper limit on the source's flux density. This approach provides a more conservative constraint than using the r.m.s. noise level of the map, scaled by the square root of the number of beams covered by the source\footnote{this would result in $\alpha <-2.5$}. Given its steep spectrum and relatively compact size, the source could represent remnant AGN plasma. However, the absence of a clear optical counterpart renders this interpretation somewhat speculative. One difficulty is that the parent object may have moved significantly across the sky, complicating identification. The possibility that the feature is a buoyant bubble of old AGN plasma that has risen up through the ICM seems less likely, as its elongation is in the wrong direction.

\begin{figure*}
\centering
\includegraphics[width=0.47\textwidth]{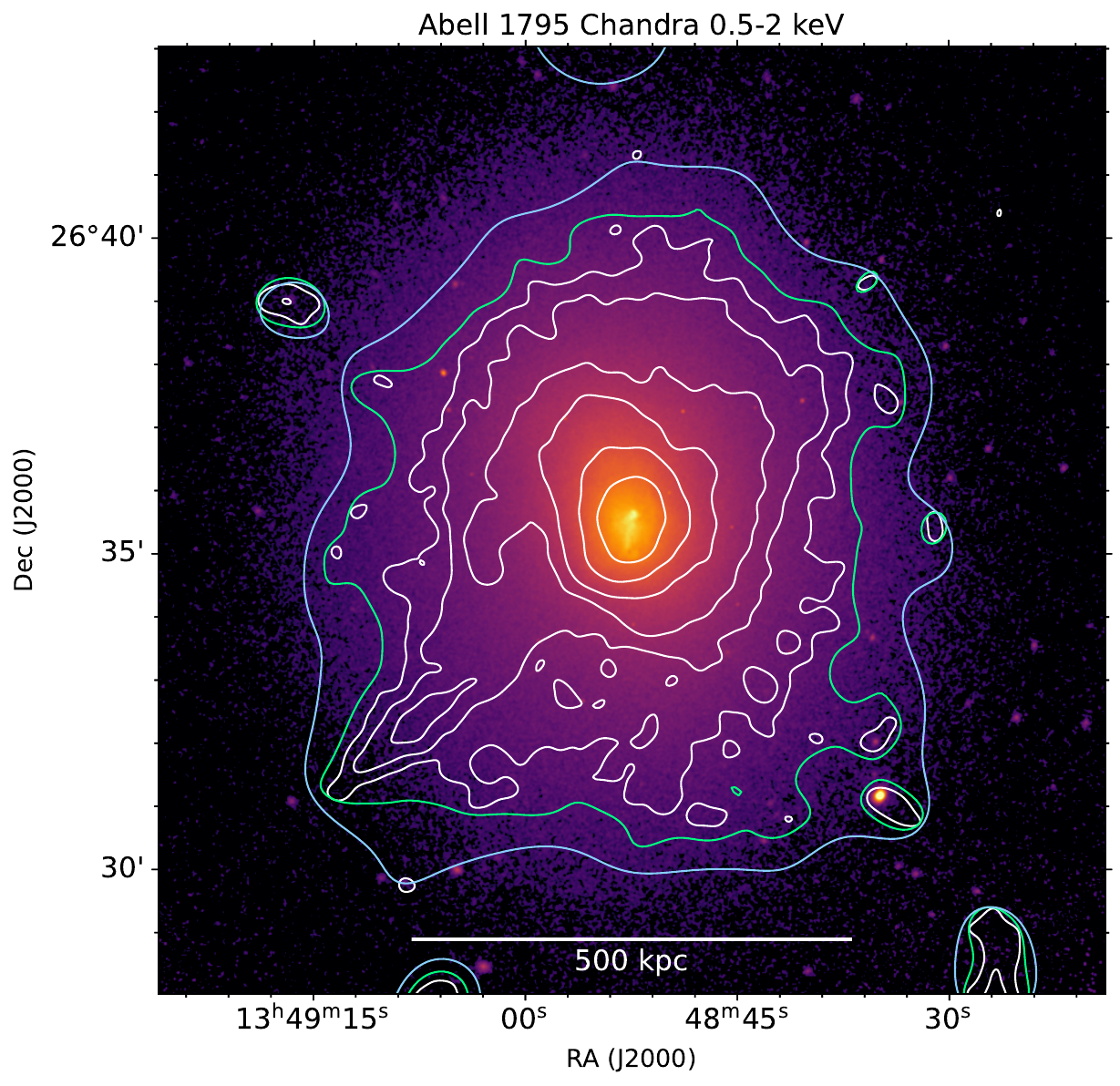}
\includegraphics[width=0.50\textwidth]{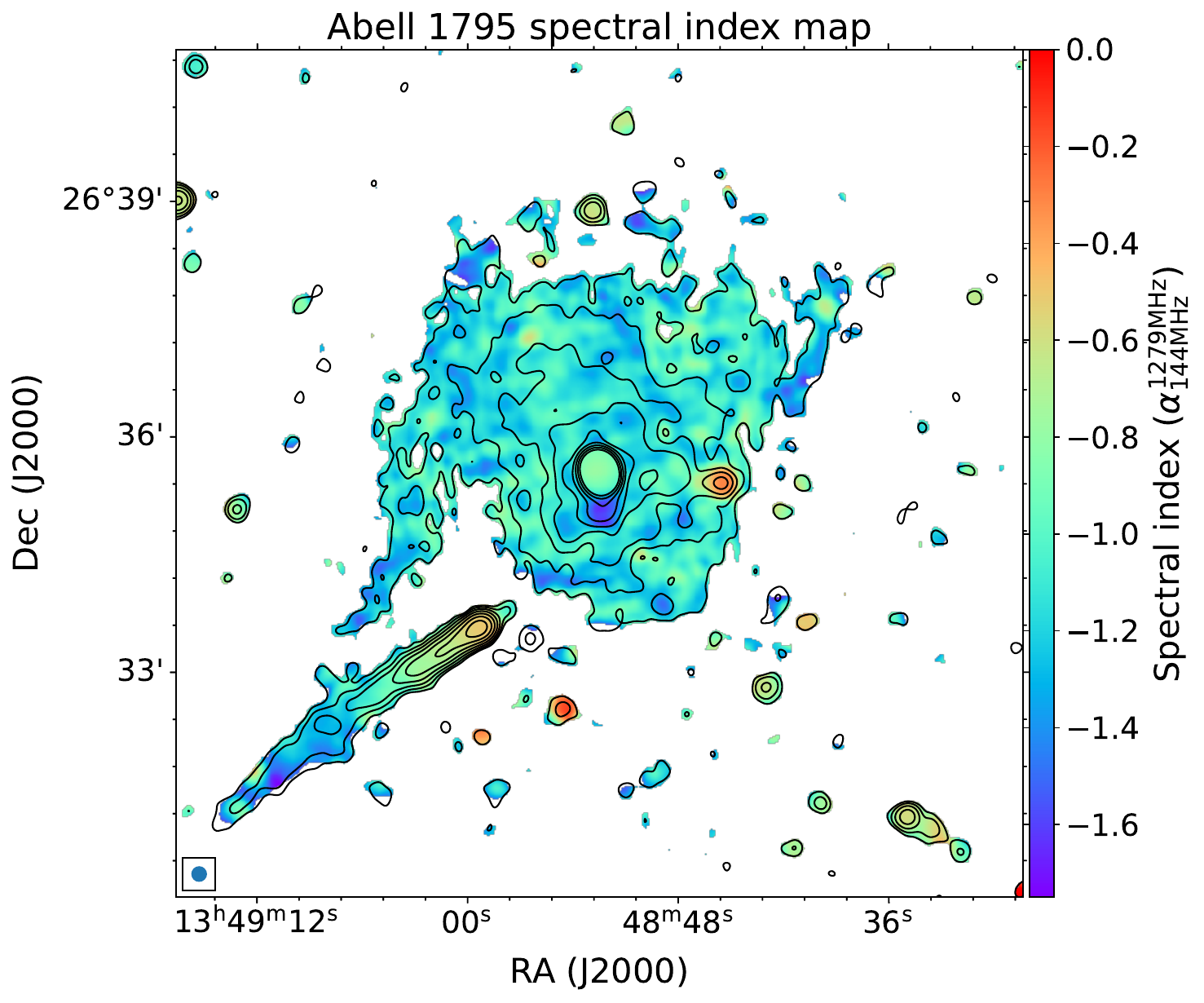}
\caption{Left panel: Chandra 0.5--2.0~keV X-ray image of Abell~1795. MeerKAT L-band contours, with emission from compact sources subtracted, are overlaid. The lowest two contour levels come from images tapered at 100~kpc (blue) resolution and 50~kpc resolution (green), see also Fig.~\ref{fig:lowres}. These are drawn at a level of $5\times$ the r.m.s. map noise ($\sigma_{\rm{rms}}$).
The white contours come from the 25~kpc resolution image and are drawn at levels of $[1,2,4,\ldots] \times 5\sigma_{\rm{rms}}$. The noise levels and beam sizes of the radio images are reported in Table~\ref{tab:imageproperties}. Right panel: Spectral index map at 12\arcsec~resolution between 144 and 1279 MHz for Abell~1795. Contours are from the 144 MHz LOFAR image and are drawn at levels of $[1,2,4,\ldots] \times 3\sigma_{\rm{rms}}$, with $\sigma_{\rm{rms}}=153$~$\mu$Jy~beam$^{-1}$. The corresponding spectral index uncertainty map is shown in Fig.~\ref{fig:spixerror}.}
\label{fig:ChandraMeerKATA1795}
\end{figure*}
\begin{figure*}
\centering
\includegraphics[width=0.49\textwidth]{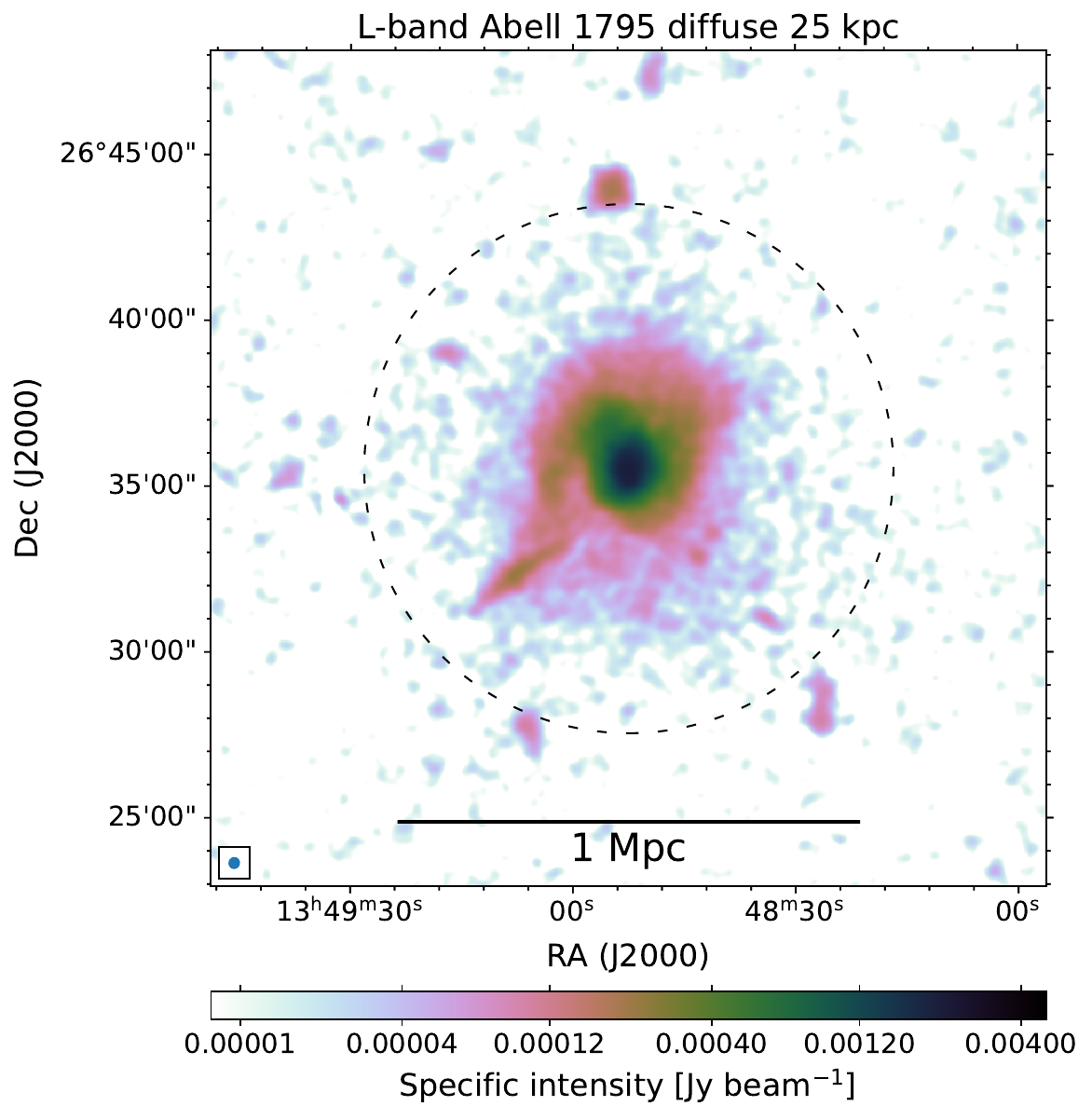}
\includegraphics[width=0.49\textwidth]{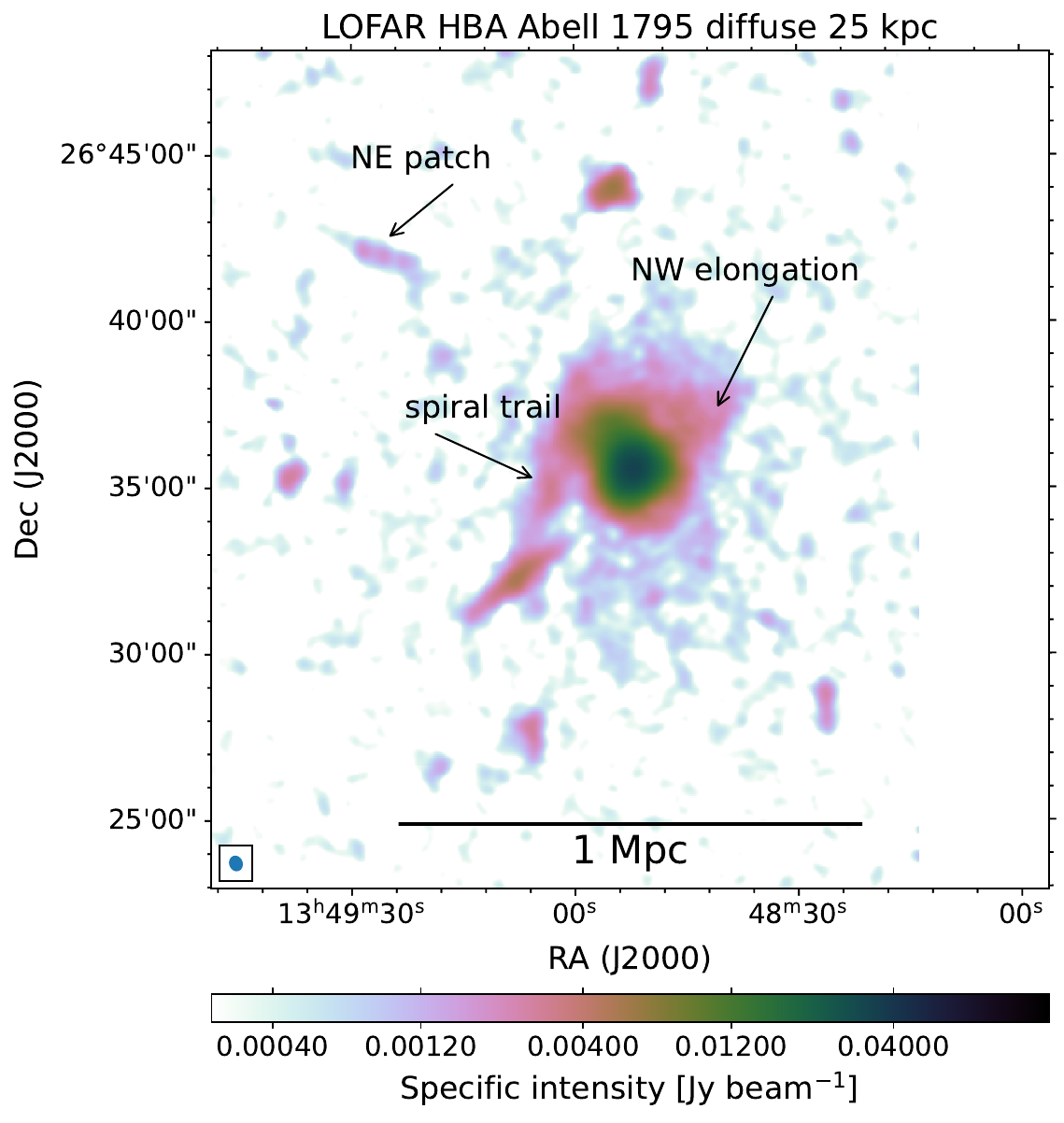}
\caption{MeerKAT (left) and LOFAR (right) images of Abell~1795 with the emission from compact sources subtracted. Both images are tapered to a resolution of 25~kpc at the cluster's redshift. The blanked area on the right side of the LOFAR image corresponds to the extraction boundary from the LoTSS pointings. Various features are labeled on the LOFAR image.  The dashed circle in the left panel indicates $0.5\times R_{500}$ \citep{2016A&A...594A..27P}. The noise levels and beam sizes of the radio images are reported in Table~\ref{tab:imageproperties}.}
\label{fig:radiodiffuseA1795}
\end{figure*}
\begin{figure}
\centering
\includegraphics[width=0.49\textwidth]{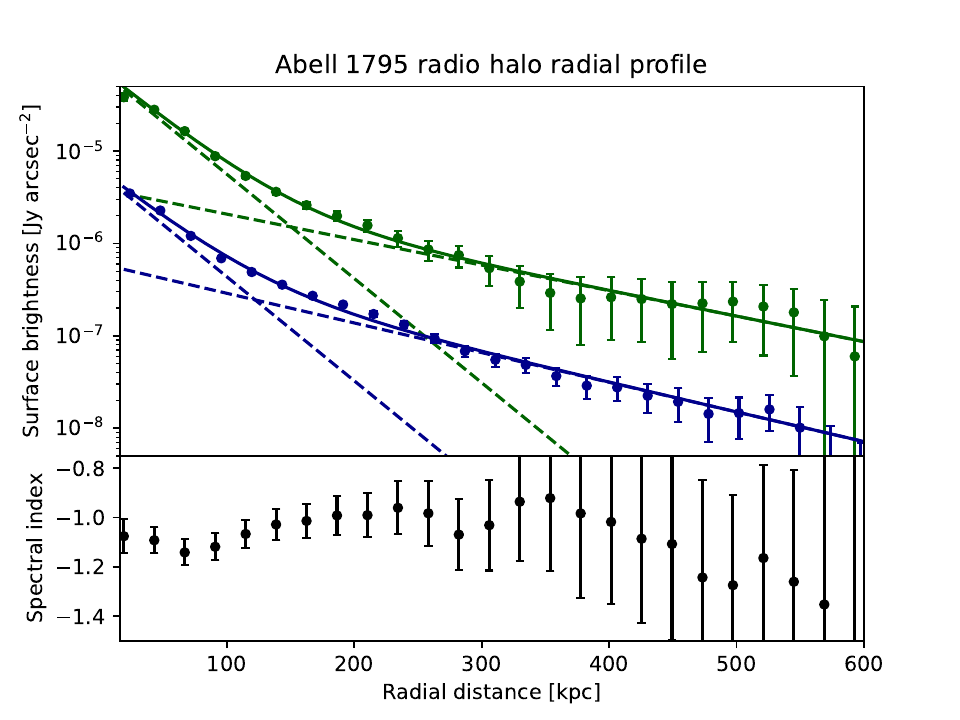}
\caption{Top panel: Radial radio surface brightness profiles for Abell~1795 extracted from the MeerKAT (blue) and LOFAR (green) 25~kpc resolution images. The solid lines show the best-fitting double exponential models (Eq.~\ref{eq:profile}). The dashed lines show the two individual exponential components. Bottom panel: The 144--1279~MHz spectral index profile computed from the data points shown in the top panel.}
\label{fig:profileA1795}
\end{figure}

The \textit{Chandra} X-ray image of the cluster from \cite{2022A&A...668A..65T}, overlaid with MeerKAT radio contours, is shown in Fig.~\ref{fig:ChandraMeerKATA1795}. This image reveals that the radio halo has a total extent of approximately 1~Mpc (see also Fig.~\ref{fig:lowres}). Similar to Abell~1775, we extract a radial radio surface brightness profile for the cluster, with the profile centred on the BCG. This profile is displayed in Fig.~\ref{fig:profileA1795}, with the masked areas shown in Fig.~\ref{fig:lowres}. Both the MeerKAT and LOFAR radio profiles reveal the presence of two distinct components. Fitting Eq~\ref{eq:profile}, we find that the data are well described by a two-component exponential model. From the best-fitting model parameters (see Table~\ref{tab:profile}), we obtain integrated flux densities of $S_{144} = 916 \pm 109$~mJy and $S_{1279} = 86.7 \pm 4.7$~mJy (integrating the model to $r = \infty$). Taking these values, we compute an integrated spectral index of $\alpha = -1.08 \pm 0.06$. For the total radio power, taking 80\% of the integrated flux density values, we derive $P_{\rm{144\mbox{ }MHz}} = \left(6.5 \pm 0.8 \right) \times 10^{24}\text{ W}\text{ Hz}^{-1}$ and $P_{\rm{1.4\mbox{ }GHz}} = \left(5.8 \pm 0.3 \right) \times 10^{23}\text{ W}\text{ Hz}^{-1}$, where we scale with $\alpha = -1.08$.

\interfootnotelinepenalty=10000

Using the radio surface brightness profiles, we derived a spectral index profile to examine its variation as a function of cluster-centric distance, see Fig.~\ref{fig:profileA1795}. The profile remains relatively flat to approximately 400~kpc within uncertainties, with typical values of $\alpha = -1.1 \pm 0.1$. This result is consistent with the radial radio surface brightness fits, which yielded similar e-folding radii for both the inner and outer halo components in the LOFAR and MeerKAT images, within the uncertainties.\footnote{The fact that the independently derived profiles from LOFAR and MeerKAT yield consistent e-folding radii within the uncertainties—despite the large difference in the number of compact sources removed—combined with the observation that the transition in these profiles occurs above the background noise level, provides  evidence that the profile slope change is not caused by residual compact-source contamination \citep{2025arXiv250505415R}.} The relatively constant distribution of the spectral index without clear trends is also apparent in the spectral index map shown in Fig.~\ref{fig:ChandraMeerKATA1795} (right panel).

\section{Discussion}
\label{sec:discussion}

\subsection{AGN feedback in Abell~1795: A ghost cavity filled with low-frequency radio emission}
\label{sec:ghost}

X-ray observations reveal a ${\sim}45$~kpc filament of cooler gas extending southward from Abell~1795's central BCG \citep{2001MNRAS.321L..33F}. This structure coincides with a luminous optical line-emitting filament first identified by \cite{1983ApJ...272...29C}. Near the southern end of the filament lies a small X-ray cavity, approximately 18~kpc in diameter\footnote{\cite{2014MNRAS.445.3444W} report a radius of 3.5~kpc, but \cite{2018A&A...618A.152K} identify a continuation of the depression extending northward.}, which was previously interpreted as a possible ``ghost cavity'' of AGN origin. However, this interpretation was considered unlikely due to the cavity’s small size and the absence of associated low-frequency radio emission \citep{2005MNRAS.361...17C,2014MNRAS.445.3444W,2018A&A...618A.152K}.

Our discovery of steep-spectrum radio emission at the cavity’s location -- emission that is physically connected to the central radio source {4C\,+26.42} -- provides strong evidence that this feature is indeed a ghost cavity filled with aged, low-frequency radio plasma originating from the central AGN. In addition, \cite{2018A&A...618A.152K} showed that the X-ray depression extends farther north. The LOFAR image reveals that the radio emission traces this extended depression and continues northward, linking it directly to {4C\,+26.42} (see Fig.~\ref{fig:A1795ghost}).

As discussed by \cite{2018A&A...618A.152K}, several scenarios have been proposed to explain the origin of the 45~kpc X-ray filament. One possibility is that it formed through gas cooling in the gravitational wake of the central galaxy as it oscillates within the cluster core \citep{2001MNRAS.321L..33F}. A direct connection with AGN activity has generally been considered unlikely \citep{2005MNRAS.361...17C,2018A&A...618A.152K}, given the absence of clear AGN-related features -- such as clear X-ray cavities or radio emission -- along the full length of the filament. Our results challenge that interpretation. The observed radio emission extends along the full length of the X-ray filament, suggesting a close physical link between the filament and AGN activity. This alignment favors a scenario in which AGN-driven outflows uplift low-entropy gas from the cluster centre to larger radii \citep[e.g.,][]{2017ApJ...844..122F}. Once uplifted, this gas then condenses into cold filaments through thermal instabilities \citep[e.g.,][]{2012MNRAS.419.3319M,2016ApJ...830...79M,2017ApJ...845...80V}. 

\subsection{Radio haloes at low frequencies and contamination from ultra-steep spectrum emission}
\label{sec:ussf}
The MeerKAT and LOFAR observations of Abell~1775 reveal strikingly different views of the cluster over a decade in frequency. In the 144~MHz LOFAR image, the emission in the cluster's core is dominated by extended ultra-steep spectrum filaments F1 and F2. In contrast, in the 1279~MHz MeerKAT image, the radio halo is the dominant smoother extended emission feature, with the filaments almost invisible. This finding is consistent with uGMRT Band~3 and 4 observations presented by \cite{2025arXiv250204913B} of the cluster, where the filaments are still clearly visible in the 400~MHz (Band 3) image, but nearly absent in the 650~MHz (Band 4) image.

Filaments F1 and F2 are likely of AGN origin, possibly involving the re-energization or revival of aged AGN plasma \citep{2002MNRAS.331.1011E,2017SciA....3E1634D,2017NatAs...1E...5V,2022ApJ...935..168R}. This interpretation is supported by their high surface brightness, comparable to that of the large-tailed AGN B1339+266B, and by the apparent morphological connection between F1 and this AGN, as well as the presence of fainter filaments linking F2 back to F1. Nevertheless, we cannot rule out the possibility that some of the fainter, steep-spectrum filaments may be associated with the radio halo. In fact, turbulence in the ICM is expected to be inhomogeneous, which can naturally generate filamentary structures \citep[see, e.g., fig.~8 in][]{2021ApJ...907...32B}, and high-resolution radio halo images have revealed that these sources are not entirely smooth \citep[e.g.,][]{2023A&A...674A..53B,2024A&A...692A..12V}.

The example of Abell~1775 highlights the need for caution when deriving integrated radio spectra of haloes in the absence of high-fidelity images and spectral index maps. If ultra-steep-spectrum emission associated with AGN activity is not properly identified, the resulting radio halo spectrum may appear artificially steep. This issue is particularly relevant for LOFAR Low Band Antenna (LBA) observations, given their lower observing frequencies and coarser spatial resolution which makes source subtraction more challenging \cite{2024A&A...689A.218P}. Moreover, recent studies \citep{2024NatAs...8..786G,2025A&A...693A..99G} indicate that filamentary ultra-steep-spectrum emission may become increasingly prevalent at very low frequencies. Consequently, robust studies of radio haloes in the LBA band require deep, high-quality, multi-frequency data to properly assess and mitigate such contamination.

\subsection{Radio (mini-)halo edges and multiple component haloes}
Recent deep observations have revealed that the emission from haloes and mini-haloes is not fully smooth but shows considerable substructure, including filaments and edges \citep[e.g.,][]{2017MNRAS.469.3872G,2022SciA....8.7623B,2023A&A...674A..53B,2024A&A...692A..12V,2025arXiv250416158B}, with some of these being co-located with X-ray surface brightness discontinuities, indicating a close link between the dynamics of the thermal and non-thermal components of the ICM. The presence of double-component radio haloes in clusters with well-defined cores and cold fronts, either sloshing or merger-induced, also seems a more general feature, with the number of examples increasing \cite[e.g.,][]{2018MNRAS.478.2234S,2019A&A...622A..24S,2021MNRAS.508.3995B,2022MNRAS.512.4210R,2024A&A...686A..82B,2024A&A...686A..44R,2024A&A...692A..12V,2024A&A...683A.132L,2024ApJ...961..133G,2025arXiv250209472H}.

In the case of Abell~1775, we observe an asymmetric radio halo brightness distribution in the cluster core with the radio emission following the drop in X-ray surface brightness at the ``leading edge'' cold front. This situation is similar to, for example, that observed in the Bullet Cluster, where the radio emission also drops at the ``bullet cold front'' as noted by \cite{2023A&A...674A..53B}. As for the Bullet Cluster, the cold front in Abell~1775 is thought to be associated with a remnant core originating from a merger event \citep{2021A&A...649A..37B}. For Abell~1795, we find that the radio emission traces the X-ray surface brightness decline at the cold front south of the cluster centre \cite{2001ApJ...562L.153M}. These cold fronts in relaxed cool-core clusters have been explained by the sloshing of gas from a minor merger event \citep{2006ApJ...650..102A}. In some cases, the morphology of mini-haloes follows these spiral-like patterns observed in the ICM \citep[e.g.,][]{2008ApJ...675L...9M,
2014ApJ...795...73G,2024MNRAS.531.4060K}.

For both Abell~1775 and Abell~1795, the radio halo emission is not fully confined within the cold fronts but extends beyond them. However, the surface brightness declines sharply across the fronts, making this extended emission more difficult to detect. Similar cases have been reported in other studies \citep[e.g.,][]{2022MNRAS.512.4210R}, suggesting that the apparent confinement of radio emission by cold fronts in some clusters may be partly due to observational limitations, with the caveat of unknown projection effects.  Simulations of mini-haloes generated by core sloshing also predict faint radio emission extending beyond the cold fronts \citep{2013ApJ...762...78Z, 2015ApJ...801..146Z}. More recently, \citet{2024ApJ...961..133G} found that the radio emission can also be bounded by more distant, outer cold fronts rather than just the inner ones, indicating that the diffuse emission may be shaped by large-scale sloshing throughout the cluster core.

A striking feature in the radio image of Abell~1795 is a spiral-like structure that extends beyond the cluster core (see the labels on the right panel of Fig.~\ref{fig:radiodiffuseA1795}), although no corresponding large-scale sloshing spiral is evident in the X-rays. Interestingly, this radio feature appears connected to the tailed radio galaxy {FIRST\,J134859.3+263334}. A similar connection has been observed in the Perseus cluster, where the radio halo seems linked to the tailed radio source IC310 \citep{2024A&A...692A..12V}. One possibility is that remnant fossil plasma from the tail is being re-accelerated by ICM turbulence. However, for this scenario to be viable, the galaxy must not be on its first infall into the cluster. Otherwise, the orientation of the spiral-like radio feature relative to the AGN tail would be difficult to explain, as the tail points away from the cluster centre. Another cluster showing a spiral-like radio morphology is RXC~J0232.2$-$4420 \citep[][]{2022A&A...657A..56K,2022MNRAS.514.5969K}, suggesting that such patterns may be quite common.

Interestingly, simulations by \cite{2021ApJ...914...73Z} reveal several cases of cosmic-ray spiral-like patterns (see their fig.~10) on a similar large scale as observed in Abell~1795. These structures arise from relativistic electrons originating from the central AGN, which are then transported by gas-sloshing motions triggered by an off-axis minor merger. Such motions can advect cosmic-ray bubbles to large radii over several gigayears. If these cosmic rays serve as seed particles for turbulent re-acceleration, they could generate features similar to those seen in Abell~1795. 

\subsection{Surface brightness point-to-point correlation}

In many radio haloes, the synchrotron emission closely follows the morphology and extent of the X-ray emission, highlighting a link between the thermal and non-thermal components of the ICM. To explore this connection, we examine the point-to-point correlation between the radio and X-ray surface brightness, following approaches used in previous studies \citep[e.g.,][]{2001A&A...369..441G,2021A&A...646A.135R,2022MNRAS.512.4210R,2023MNRAS.524.6052R}.

The slope of this correlation provides insight into the underlying physical processes. Assuming that CR protons are diffused on cluster scales, in hadronic (secondary) models, a super-linear slope is generally expected due to the centrally peaked distribution of cosmic ray protons (CRp) and their scaling with the thermal gas \citep[see][]{2020A&A...640A..37I}. In contrast, turbulent re-acceleration models can yield either sub-linear or super-linear slopes, depending on the distribution and properties of the relativistic electrons (CRe) throughout the cluster volume.

We characterise this relationship by fitting a power-law in log-log space of the form
\begin{equation}
\log(I_R) = c + k \log(I_X),
\label{eq:p2p}
\end{equation}
where the slope $k$ describes how the non-thermal emission scales with the thermal emission.

For each cluster, we performed the following steps. We smoothed the \textit{Chandra} X-ray maps to match the 25~kpc spatial resolution of the MeerKAT data (corresponding to angular resolutions of 22\arcsec{} for Abell~1795 and 20\arcsec{} for Abell~1775). Regions affected by discrete sources and contaminating radio emission were masked in both the MeerKAT and \textit{Chandra} maps. We then covered the area of ``significant'' X-ray emission—approximately above the $3\sigma$ level—with boxes at the same spatial resolution.

We extracted flux density measurements for all selected regions and used \texttt{Linmix} \citep{2007ApJ...665.1489K} to quantify the point-to-point correlation. \texttt{Linmix} applies a Bayesian linear regression approach that incorporates uncertainties in both variables, intrinsic scatter, and importantly, upper limits on the dependent variable ($I_R$). We employed its MCMC implementation to derive posterior distributions for the fit parameters. Uncertainties on the slope and intrinsic scatter are quoted at the $1\sigma$ level. The strength of the correlation was further characterised by calculating the Spearman ($r_S$) and Pearson ($r_P$) correlation coefficients. In our analysis, regions where the radio surface brightness was below $2\sigma$ were treated as $2\sigma$ upper limits.

\begin{figure*}
\centering
\includegraphics[width=0.49\textwidth]{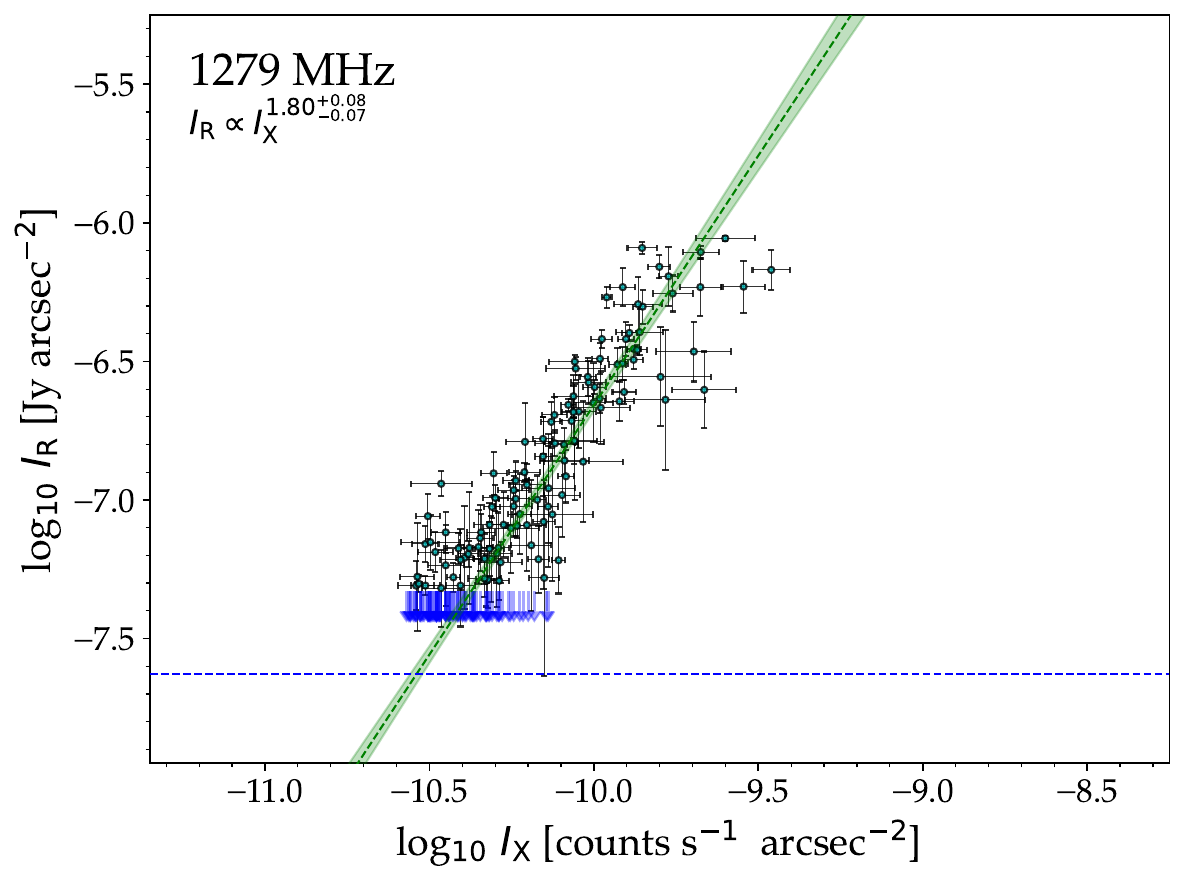}
\includegraphics[width=0.49\textwidth]{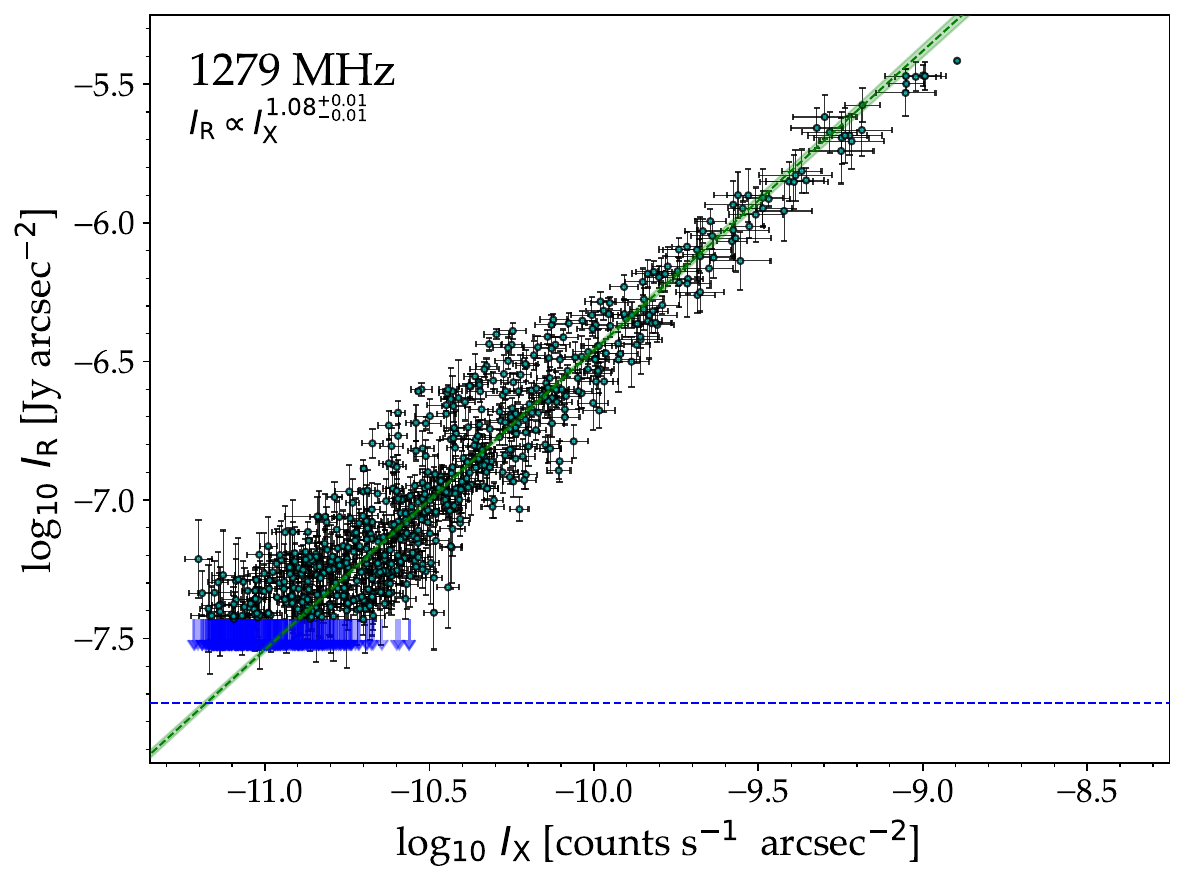}
\caption{Point-to-point correlation between radio and X-ray surface brightness. The left panel shows the $I_R$--$I_X$ relation for Abell~1775 based on the MeerKAT and Chandra data, while the right panel displays the same analysis for Abell~1795. Regions where the radio surface brightness falls below the $2\sigma$ level are treated as $2\sigma$ upper limits and are marked with blue arrows; the dashed blue line indicates the $1\sigma$ threshold. The dashed green line represents the best-fit power-law relation, and the shaded green area denotes the corresponding $1\sigma$ uncertainty.}
\label{fig:PTP}
\end{figure*}

The resulting best-fit slope parameters ($k$) and intrinsic scatter and the correlation coefficients are summarised in Table~\ref{tab:ptp}. The point-to-point correlations themselves are shown in Fig.~\ref{fig:PTP}.

\begin{table}
    \centering
    \caption{Results of the point-to-point correlation analysis.}
    \label{tab:ptp}
    \begin{tabular}{lcc}
        \hline\hline
        & Abell~1795 & Abell~1775 \\
        \hline
        Slope $k$ & $1.080^{+0.014}_{-0.014}$ & $1.800^{+0.077}_{-0.074}$ \\
        Intrinsic scatter & $0.020^{+0.002}_{-0.002}$ & $0.025^{+0.006}_{-0.005}$ \\
        Spearman coeff. $r_S$ & $0.92$ & $0.90$ \\
        Pearson coeff. $r_P$ & $0.95$ & $0.90$ \\
        \hline
    \end{tabular}
\end{table}

Overall, both clusters show strong, tight correlations between the radio and X-ray surface brightness. Abell~1775 has a distinctly super-linear slope, typical of a mini-halo\footnote{The outer component's radio emission is very faint so we cannot determine the slope for this component separately.}.
This is likely still consistent with the slingshot scenario proposed by \cite{2021A&A...649A..37B} as the origin of the radio emission, since both slingshot and sloshing--commonly invoked for mini-haloes--arise under similar conditions. The main difference between a classical mini-halo and a slingshot radio halo would then lie in the dynamics that drive the gas motions in the ICM.

For Abell~1795, the slope is marginally super-linear (though statistically significant). A notable aspect of these results is the absence of any clear deviation from a single power-law slope in the point-to-point (PTP) correlations, despite the radial radio profiles showing multiple distinct components. Similar hints were reported for MS~1455.0+2232 by \cite{2022MNRAS.512.4210R}. The measured slope of $k=1.08\pm0.01$ is also noteworthy--steeper than typical giant haloes in merging clusters, yet shallower than most mini-haloes. Again in this respect, Abell~1795 is comparable to MS~1455.0+2232, which exhibits one of the flattest slopes among known mini-haloes \citep{2022MNRAS.512.4210R,2020A&A...640A..37I}.

\begin{figure*}
\centering
\includegraphics[width=0.49\textwidth]{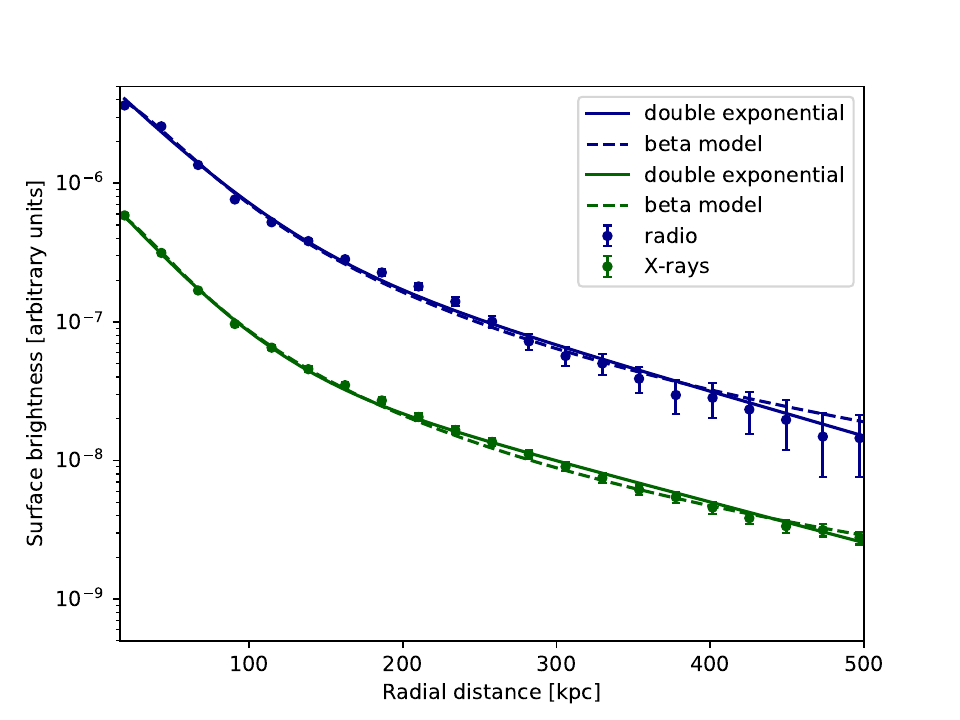}
\includegraphics[width=0.49\textwidth]{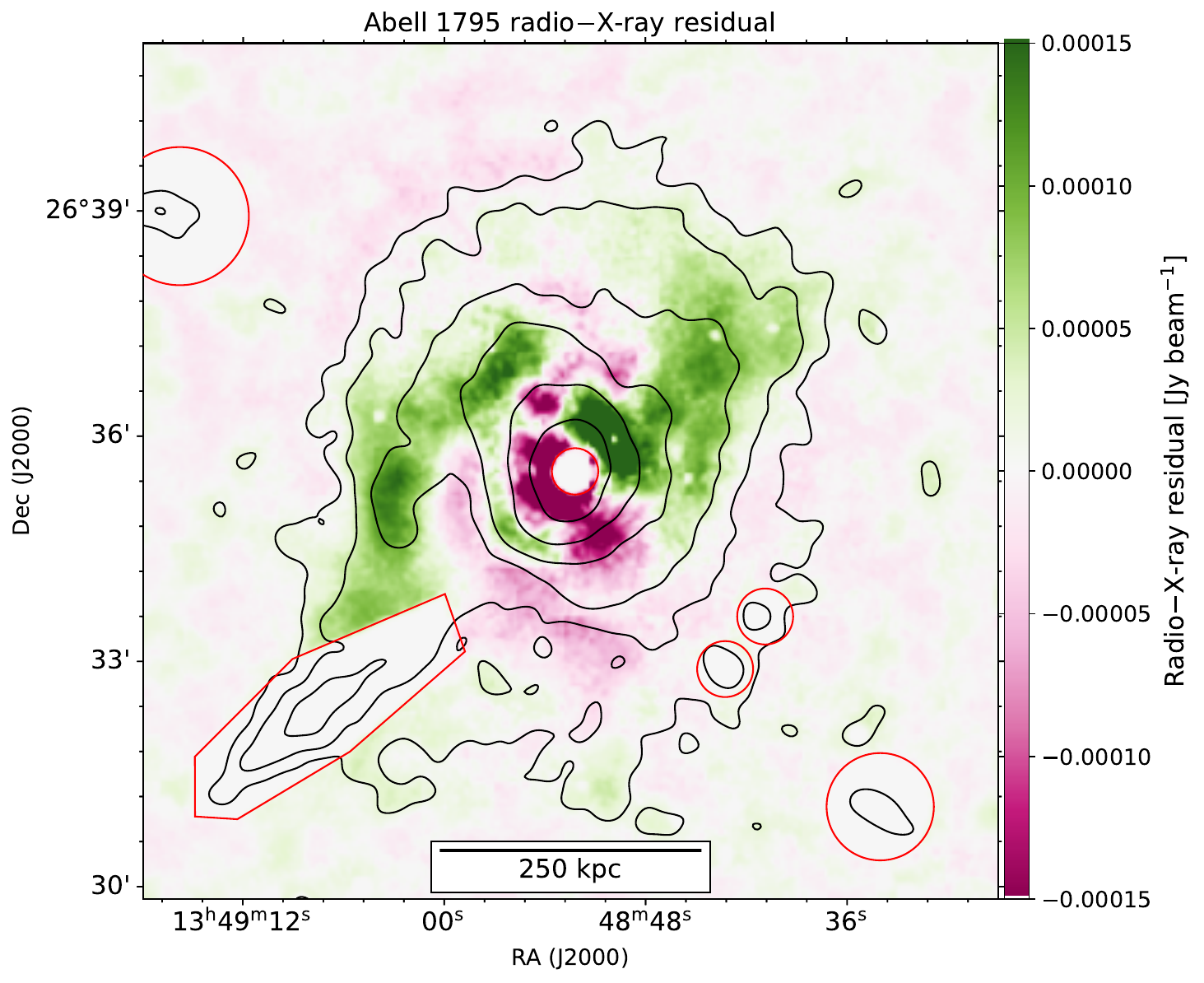}
\caption{Left: A comparison of the radio and X-ray surface brightness profiles for Abell~1795. Fits to the profiles are shown with dashed and solid lines. Right: Radio--X-ray residual (RXR) image (with the emission from compact sources removed) obtained by subtracting a scaled version of the X-ray image according to Eq.~\ref{eq:RXR}. The regions indicated by the red contours were masked. The black contours correspond to the 25~kpc resolution MeerKAT image and are identical to the white contours shown in the left panel of Fig.~\ref{fig:ChandraMeerKATA1795}.} 
\label{fig:A1795radioxrays}
\end{figure*}

For Abell~1795, the tight correlation and slope close to unity indicate that the radio emission closely follows the X-ray brightness distribution. This is further supported by the nearly identical shapes of the X-ray and radio radial profiles, see Fig.~\ref{fig:A1795radioxrays} (left panel). Consequently, instead of the double-exponential model of Eq.~\ref{eq:profile}, a $\beta$-model--commonly used to describe X-ray surface brightness profiles \citep{1976A&A....49..137C,2009A&A...500..103A}--provides a more physically motivated fit. Indeed, the $\beta$-model offers an excellent description of both the radio and X-ray profiles, as shown in Fig.~\ref{fig:A1795radioxrays}.

The close correspondence between the radio and X-ray images motivates us to search for regions of excess non-thermal radio emission relative to found point-to-point correlation with the thermal component. To do this, we compute a radio–X-ray residual (RXR) map defined as
\begin{equation}
\mathrm{RXR}(x,y) = I_R(x,y) - 10^c \cdot I_X(x,y)^{k},
\label{eq:RXR}
\end{equation}
where $10^c$ is a normalization factor (see Eq.~\ref{eq:p2p}). We adopt $k = 1.08$ and $c$ from the point-to-point fitting  to produce the RXR map. The resulting map is shown in the right panel of Fig.~\ref{fig:A1795radioxrays}. The RXR map reveals two interesting regions. First, there is an excess of radio emission associated with the spiral-like trail extending toward the tailed radio galaxy \textit{FIRST,J134859.3+263334}. Second, an excess of emission is observed extending northwest of the cluster centre, corresponding to the ``NW elongation'' marked in Fig.~\ref{fig:radiodiffuseA1795}.

We note that RXR maps provide an efficient way to identify deviations from the point-to-point radio–X-ray correlation slope. While previous studies have typically examined selected regions individually to assess spatial variations in the correlation, RXR maps offer a more systematic method for locating and analyzing such deviations. This approach might be preferable to selecting regions based solely on features seen in the radio maps.

\subsection{Origin of the Abell~1795 radio halo}

The striking similarity between the Abell~1795 radio and X-ray morphologies suggests a single underlying physical mechanism for both the inner and outer radio components. The two radio components appear to mirror structures already present in the thermal ICM, implying that the entire radio structure could be regarded as a single, large-scale radio halo. In this view, distinguishing between mini-halo and giant halo components may not be meaningful.

For giant radio haloes in merging clusters, there is a well-established statistical correlation between cluster mass and radio power, as well as with various mass proxies such as the X-ray luminosity and the integrated Compton $Y$ parameter \citep[e.g.,][]{2000ApJ...544..686L,2008A&A...480..687C,2012MNRAS.421L.112B,2013ApJ...777..141C,2015A&A...579A..92K}. This correlation is thought to arise because a fraction of the gravitational energy released during cluster mergers is converted into magnetic field amplification and the re-acceleration of cosmic rays through turbulence in the ICM \citep[e.g.,][]{2005MNRAS.357.1313C,2023A&A...672A..43C}. Within this framework, minor mergers may also generate radio haloes; however, such cases are expected to be less common, and the resulting haloes are statistically predicted to have steeper spectral indices \citep[e.g.,][]{2006MNRAS.369.1577C}.

Recent studies have reported extended, cluster-scale radio emission in some more relaxed clusters \citep[e.g.,][]{2017A&A...603A.125V,2018MNRAS.478.2234S,2023A&A...678A.133B,2024A&A...686A..82B,2024A&A...686A..44R}. In the case of the Perseus cluster, for example, the radio power is consistent with the mass–power relation, and turbulence generated by an off-axis merger has been proposed as a sufficient mechanism to power the observed Mpc-scale radio emission \citep{2024A&A...692A..12V}.

In the turbulent re-acceleration scenario, a steeper spectral index beyond the cool-core region is expected if the available turbulent energy is lower than within the sloshing core \citep[e.g.,][]{2005MNRAS.357.1313C,2008Natur.455..944B}. For predominantly relaxed clusters, the level of turbulence outside the core is likely lower than in merging systems. Consequently, giant radio haloes in cool-core clusters are expected to exhibit steeper spectra than those in dynamically disturbed clusters. This aligns with the observation that radio haloes discovered with LOFAR are more frequently found in less disturbed systems \citep{2023A&A...672A..43C}, which would be expected if these haloes have steeper spectra and are thus more easily detected at lower frequencies.

In Fig.~\ref{fig:m_p150}, we plot the location of the Abell~1795 radio halo (both components) on the $\rm{M}_{500} -P_{\rm{150~MHz}}$ relation, using data from \cite{2022A&A...660A..78B} and overlaying the best-fitting correlation from \cite{2023A&A...680A..30C}. We find that the radio halo in Abell~1795 aligns well with this correlation. This conclusion holds regardless of whether the inner component is included, as both components contribute approximately equally to the total power, placing the system comfortably within the correlation's scatter. In fact, removing the inner component shifts the halo even closer to the correlation. Furthermore, the central surface brightness $I_0$ and the 144~MHz characteristic radius $r_{\rm e} = 157 \pm 35$~kpc of the giant halo in Abell~1795 are consistent with those of the giant radio haloes studied by \cite{2022A&A...660A..78B}. Although $r_{\rm e}$ is below the sample mean of 194~kpc, it lies within the standard deviation of 94~kpc reported in \cite{2023A&A...672A..41B}. In terms of its spectral index, the radio halo also resembles the properties of known haloes in massive (merging) clusters \citep[e.g.,][]{2021A&A...654A..41R,2022ApJ...933..218B}. In this sense, the spectral, size, and total power properties of the Abell~1795 halo are similar to those of other known giant radio haloes. 

The properties of the Abell~1795 halo thus appear as an outlier among the population of haloes predicted by the turbulent acceleration scenario. One would anticipate either a steeper spectrum or the absence of a giant radio halo at frequencies above a gigahertz in a relaxed cluster with mass $M_{500}=\left(4.47\pm0.14\right) \times 10^{14}$~M$_{\odot}$ \citep{2016A&A...594A..27P}. Invoking a scenario where the cluster is somehow significantly more dynamically disturbed than other cool-core clusters is not supported by X-ray observations. 
Abell~1795 is classified as a strong cool-core cluster \citep[e.g.,][]{2010A&A...513A..37H}. While the cluster is not entirely relaxed -- there is evidence of sloshing -- such features are common in systems that are not undergoing major mergers. The presence of a single dominant BCG, a strong cool core, and largely symmetric large-scale X-ray emission all indicate that this cluster is not undergoing a major (off-axis) merger. The cluster's relaxed dynamical state is further supported by its X-ray concentration parameter ($c = 0.4260 \pm 0.00177$) and centroid shift ($w = 0.00867 \pm 0.00243$), as reported by \cite{2022A&A...660A..78B}. These values place the cluster in the ``relaxed quadrant'' according to the classification scheme of \cite{2010ApJ...721L..82C}.

Although unusual, the presence of a giant radio halo in a relaxed cluster is not inconsistent with the turbulent re-acceleration model, provided that such haloes occur predominantly during mergers, giving rise to the observed radio bimodality \citep[e.g.,][]{2007ApJ...670L...5B,2013ApJ...777..141C,2023A&A...672A..43C,2023A&A...680A..30C}.
Timescales may also be important: a recent injection of turbulence could yield a flatter spectral index even if the total turbulent energy is low. Similarly, a minor merger could produce a comparable turbulent energy density to a major merger if the energy is deposited in a smaller volume. Thus, it is interesting to note that the relatively small characteristic radius, $r_{\rm e}$, compared to the average giant halo population, suggests that the Abell~1795 radio halo is roughly two times less voluminous than a typical halo.

\subsubsection{A hadronic origin?}
\label{sec:hardonic}
It is remarkable that the radio brightness profile closely mirrors the X-ray profile, indicating that radial variations in radio emission closely track those in the X-rays. This correspondence may point to a significant contribution from secondary electrons produced through hadronic collisions with thermal protons \citep[e.g.,][]{2004A&A...413...17P,1999APh....12..169B,2024MNRAS.527.1194K}. In such a case, the radio–X-ray connection would naturally arise from the radial distribution of the thermal targets. A hadronic origin could also account for the relatively flat integrated radio spectrum. However, the nearly linear slope of $k=1.08$ between radio and X-ray brightness is not straightforward to reproduce in hadronic models. It would require a rapid increase of the ratio between the magnetic (or cosmic rays) to thermal energy densities with distance \citep[e.g.,][]{2001A&A...369..441G}

To explore these issues, we quantitatively model the radio halo surface brightness profile of Abell~1795 under the assumption that it originates from hadronic interactions between cosmic-ray protons and thermal ions in the ICM, following the approach of \cite{2024A&A...688A.175O} and references therein. To describe the non-thermal component, we assume that the cosmic ray proton population follows a power-law distribution that traces the spatial distribution of the thermal gas:
\begin{equation}
N_{\mathrm{CRp}}(p_p, R) = C_p \left[ n_{\mathrm{th}}(R)  kT(R) \right]^a p_p^{-s},
\label{eq:CRspectrum}
\end{equation}
where $n_{th}(R)$ and $T(R)$  are the thermal electron density and temperature distribution as a function of radius $R$, respectively, and $k$ is the Boltzmann constant. The $C_p$ is a normalization constant, and ${p_p}^{-s}$  represents a power-law momentum distribution of the cosmic ray protons with index $s$. 
The parameter $a$ governs the scaling between the energy densities of cosmic ray protons and the thermal plasma, with $W_\mathrm{CRp}\propto {W_\mathrm{th}}^{a}$.

Secondary electrons and positrons, as well as gamma rays, are produced via the decay of neutral pions generated in collisions between cosmic rays and thermal protons \citep[e.g.,][]{1999APh....12..169B}. The spectrum of these secondary electrons is computed following \cite{2017MNRAS.472.1506B}. In galaxy clusters, where cosmic ray protons remain confined over Gyr timescales and proton-proton collision timescales exceed electron cooling times, the secondary electron population is expected to reach a steady-state balance between injection and radiative losses. Thus, assuming a power-law distribution of cosmic ray protons given by Eq.~\ref{eq:CRspectrum}, the spectrum of the synchrotron emissivity can be approximated with a power-law. 

Projected onto the sky at a cluster-centric distance $r$, the synchrotron emission is proportional to 
\begin{equation}\label{eq:Isynch}
{I_{\mathrm{syn}}(r)} 
\propto \int_\mathrm{LOS} \frac{RdR}{\sqrt{R^2-r^2}} n_\mathrm{th}^2(R) kT(R) \mathcal{F}(R) \frac{B^{1-\alpha}(R)}{B^2(R)+B^2_\mathrm{CMB}},
\end{equation}
where we have dropped the normalisation constants. Here we define $\mathcal{F}(R)= \frac{W_\mathrm{CRp}(R)}{W_\mathrm{th}(R)}$. The spectral index is given by $\alpha =\left(1-s\right)/2$, $B$ is the ICM magnetic field strength, and $B_\mathrm{CMB}$ the equivalent magnetic field strength of the cosmic microwave background. With the assumption of Eq.~\ref{eq:CRspectrum}, where $a$ governs the scaling between the thermal plasma and cosmic ray proton energy density, we can write
\begin{equation}
\mathcal{F} = \left(W_\mathrm{th}\right)^{a-1}.
\end{equation}

For Abell~1795, we adopt the ICM electron density $n_\mathrm{th}(R)$ and temperature $T(R)$ profiles derived by \cite{2017ApJ...843...76A} from Chandra X-ray observations. The electron density profile is provided in analytic form, while the temperature profile is derived via spline interpolation of the radial bin measurements reported in that work. For the magnetic field strength $B(R)$, we assume it scales with the thermal energy density of the gas, following a commonly adopted form \citep[e.g.,][]{2010A&A...513A..30B}:
\begin{equation}\label{eq:Bfield}
B(R) = B_0 \left( \frac{n_\mathrm{th}(R)}{n_\mathrm{th}(0)} \right)^{1/2} \propto n_\mathrm{th}(R)^{1/2},
\end{equation}
where $B_0$ denotes the central magnetic field strength. For the radio halo spectral index, we use our measured $\alpha=-1.08$.

\begin{figure*}
\centering
\includegraphics[width=0.49\textwidth]{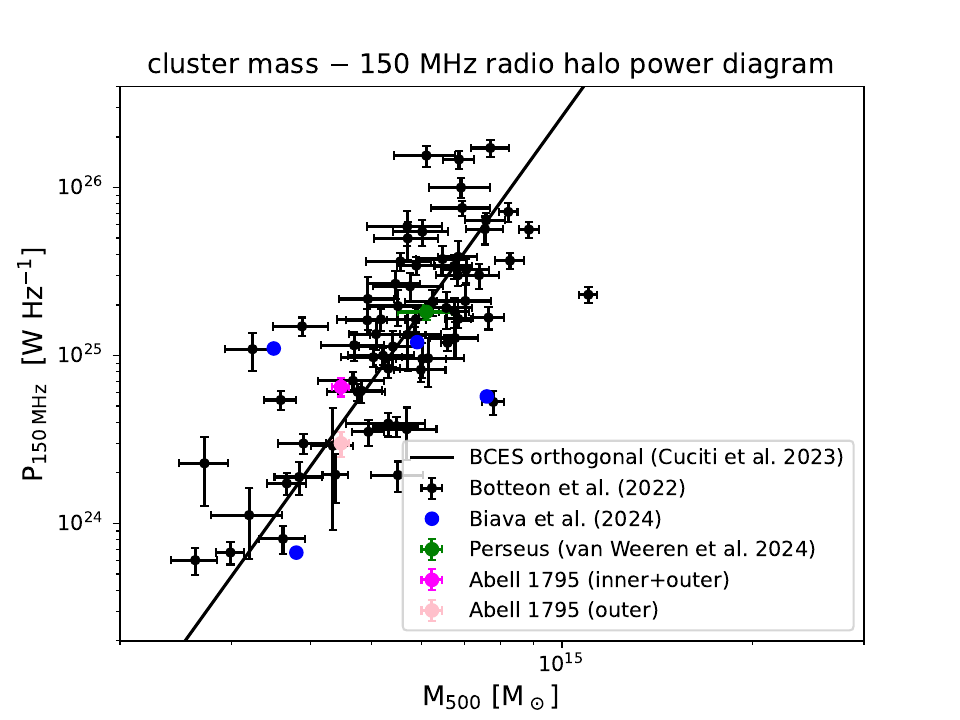}
\includegraphics[width=0.49\textwidth]{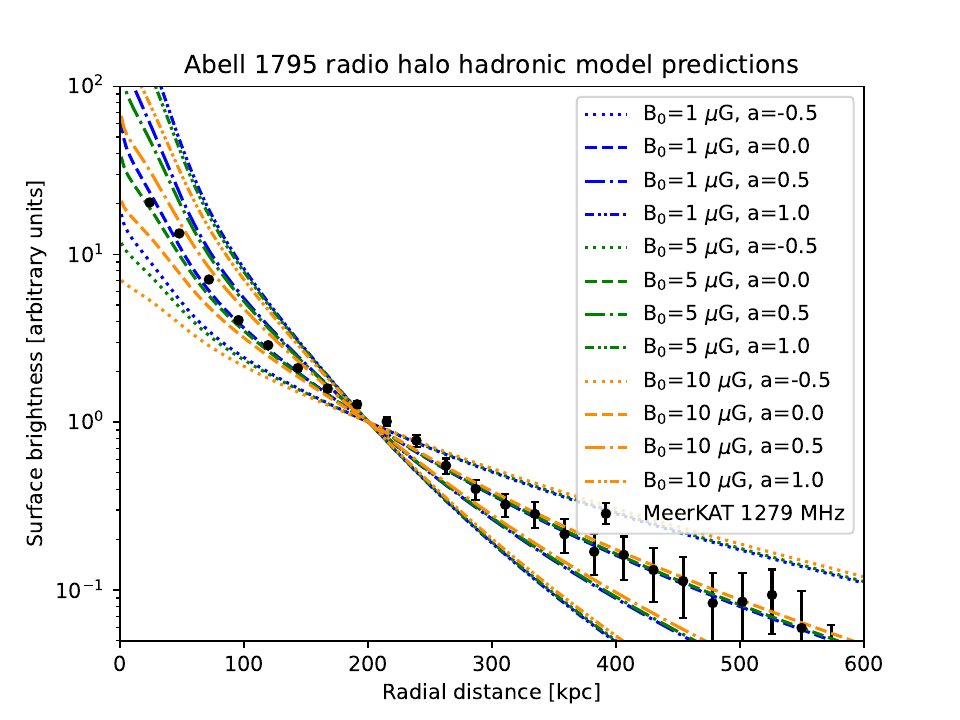}
\caption{Left panel: Cluster mass (M$_{500}$) against 150~MHz radio power. Abell~1795 is plotted in magenta (both halo components) and pink (outer component only) colours. The black data points are from \protect\cite{2022A&A...660A..78B}. The solid line shows the best-fit correlation using the BCES orthogonal method from \protect\cite{2023A&A...672A..43C}. The blue points are from \protect\cite{2024A&A...686A..82B}, where the emission from the inner-halo component was removed. The Perseus cluster halo is shown in green \citep{2024A&A...692A..12V}. Right panel: Modeled radio halo synchrotron surface brightness profile for Abell~1795 based on hadronic interactions. Data points at 1279~MHz from MeerKAT are shown in black. Profiles are computed for three values of the central cluster magnetic field strength $B_0$ and the scaling between cosmic ray energy density and thermal energy density $a$. For more details, see Sect.~\ref{sec:hardonic}. The computed and observed profiles are normalised at a radial distance of 200~kpc.}
\label{fig:m_p150}
\end{figure*}

The values of $B_0$ and the parameter $a$ (the scaling between the cosmic ray and thermal energy densities) are not known a priori. We therefore compute model radio synchrotron intensity profiles $I_{\mathrm{syn}}(r)$ 
for three $B_0$ and $a$ values, and compare these to the observed surface brightness profile. The results are shown in Fig.~\ref{fig:m_p150} (right panel).
From this comparison, we find that models with $a=0$ provide the best match to the observed radio profile, requiring that the energy density in cosmic ray protons is constant as a function of radius\footnote{This does not exclude local deviations, as the profile represents an azimuthal average.}. Models with $B_0=1$~$\mu$G and $B_0=5$~$\mu$G both reproduce the data reasonably well, indicating that the preferred central magnetic field strength lies within this range since $B_0=10$~$\mu$G yields a profile that is too flat near the cluster centre compared to observations.

It should be noted that our assumption of $B(R) \propto n_\mathrm{th}(R)^{1/2}$ affects the derived value of $a$. If this  $B(R) \propto n_\mathrm{th}(R)^{\eta}$ scaling is flatter, so $\eta<0.5$, a higher value of $a$ can reproduce the Abell~1795 radial profile. We find that, to match the observed radio profile with $a = 0.5$ or $a = 1$, the model requires $\eta \approx 0.25$ or $\eta \approx 0.0$, respectively, see Fig.~\ref{fig:otheretas}. Interestingly, a recent polarization study of a cluster sample by \cite{2025A&A...694A..44O} found that values around $\eta \approx 0$ provided a better match to the data than the canonical $\eta = 0.5$. However, earlier studies have also reported that $\eta \gtrsim 0.5$ may be appropriate \citep[e.g.,][]{2012A&A...540A..38V,2017A&A...603A.122G,2021MNRAS.502.2518S}. Future studies of the magnetic field properties of Abell~1795 will therefore be important for further testing hadronic models. However, this will be challenging, as no polarised sources have been reported in the cluster region by \cite{2022A&A...665A..71O,2025A&A...694A..44O}.

On the modeling side, it would be valuable to further explore the possible hadronic origins of the halo. In fact, the required values of $a$ and/or $\eta$ close to zero represent a challenge because the ratio between non-thermal (cosmic rays plus magnetic fields) and thermal energy must remain within reasonable limits, and the non-thermal components cannot account for an excessive fraction of the total energy budget in the external regions. Hybrid scenarios--such as the turbulent re-acceleration of secondary electrons \citep{2011MNRAS.410..127B,2017MNRAS.472.1506B}--and their impact on the resulting radio radial profiles and point-to-point correlations provide therefore additional important avenues for future investigation to test the role of secondaries in powering the Abell~1795 radio halo.

\subsubsection{Observational biases against radio haloes in cool-core systems}
The discovery of a Mpc-scale radio halo in Abell~1795, despite the existence of previous studies with the VLA, MeerKAT, GMRT, and LOFAR, provides an example which implies the presence of observational biases against detecting large radio haloes in cool-core clusters, with the Perseus Cluster providing another key example \citep[][]{2024A&A...692A..12V}. An interesting case in this context is {CL\,1821+643}, where \citet{2014MNRAS.444L..44B} reported the presence of a large radio halo in a cluster with a strong cool core. \citet{2016MNRAS.459.2940K} measured an integrated spectral index of $-1.0 \pm 0.1$, providing another example of a halo in a cool-core cluster with a relatively flat spectrum. However, unlike Abell~1795, observations suggest that CL~1821+643 experienced an off-axis merger event that left the cool core intact.

Future work is required to assess whether biases against detecting large radio haloes in cool-core clusters significantly affect the observed radio halo bimodality. This will require expanding the sample of relaxed clusters observed with deep MeerKAT and LOFAR data, while explicitly mitigating biases related to limited dynamic range. In addition, more measurements of radio halo spectral indices in cool-core clusters are essential, as the current observational constraints remain sparse. The faint nature of outer halo components demands observations with both high sensitivity to diffuse, large-scale emission and sufficient angular resolution to minimise contamination from compact sources.

Upcoming surveys with the Square Kilometre Array (SKA) will represent a major step forward in this regard. The SKA will provide excellent uv-coverage for high-dynamic-range imaging of bright compact sources, together with high sensitivity to extended emission. At the lowest SKA-Low frequencies, angular resolution may become a limiting factor; however, at frequencies of a few hundred megahertz and above, detailed spatial characterization of radio halo spectral indices out to large cluster-centric radii will be possible. This will enable robust tests of the origin of radio haloes in cool-core clusters.

\section{Conclusions}
\label{sec:conclusion}
In this paper, we presented MeerKAT L-band continuum observations of the galaxy clusters Abell~1775 and Abell~1795. Both clusters exhibit cold fronts and structured cores in X-ray observations. To complement the MeerKAT data, we also incorporated LOFAR HBA observations. We produced high-quality MeerKAT and LOFAR images for both clusters using an improved version of the \texttt{facetselfcal} procedure described in \cite{2021A&A...651A.115V}, demonstrating its improved interoperability for using it for instruments other than LOFAR.
In the case of Abell~1795, this led to a significant improvement in image quality compared to previously published MeerKAT and LOFAR images based on the same underlying data.

Below, we summarise our main scientific findings.

\begin{itemize}
    \item We detect radio haloes in both Abell~1775 and Abell~1795. In their radial surface brightness profiles, two distinct components are present. An inner component, with a higher central surface brightness and smaller characteristic radius, and an outer component with a lower central surface brightness and larger characteristic radius. The inner components can be associated with X-ray cores and also show radio surface brightness declines at the location of X-ray-detected cold fronts. 

    \item We confirm that the core of Abell~1775 is filled with filamentary ultra-steep spectrum emission, with spectral index values reaching $\alpha=-3$, consistent with the results of \cite{2021A&A...649A..37B,2025arXiv250204913B}. This filamentary emission dominates the LOFAR image but is nearly absent in the MeerKAT L-band image. This case illustrates the coexistence of multiple emission components within galaxy clusters and highlights the challenges in accurately determining the spectral index distribution of radio haloes in the presence of ultra-steep-spectrum filaments, which may either be AGN-related or physically connected to the halo itself.
    
    \item In Abell~1795, we detect steep-spectrum radio emission extending southward from the central radio source {4C\,+26.42}. This emission closely follows a 45~kpc X-ray filament of cooler gas, which is also associated with an optical line-emitting filament. The radio emission terminates at a previously identified X-ray depression. These results indicate that both the X-ray filament and the depression are connected to an outflow of plasma from the central AGN, in contrast to earlier interpretations.
    
    \item The radio halo in Abell~1795 displays a spiral-like morphology, with a trail of emission extending to a tailed radio galaxy {FIRST\,J134859.3+263334}, somewhat similar to what was found for the Perseus cluster halo and the tailed source IC~310.
    \item 
    We analyzed the point-to-point correlation between the radio and X-ray surface brightness in Abell~1775 and Abell~1795. Both clusters show strong, tight correlations. Abell~1775 displays a distinctly super-linear slope ($k=1.80\pm0.08$), typical of a mini-halo, while Abell~1795 shows a marginally super-linear slope ($k=1.08\pm0.01$), which is somewhat unusual for giant radio haloes that generally exhibit sub-linear slopes and for mini-haloes that display super-linear slopes. Despite the presence of multiple components in the radial radio profiles, the point-to-point correlations are well described by a single power law. For Abell~1775, however, the low number of data points and the large error bars in the region beyond the cluster core mean that a slope change could easily remain undetected. For Abell~1795, on the other hand, the result is significant, and the single slope might hint at a common underlying physical origin for the inner and outer halo components.

     \item The striking similarity between the radio and X-ray morphologies of Abell~1795 suggests a single underlying physical mechanism driving both the inner and outer radio components. The two radio components appear to trace structures already present in the thermal ICM, implying that the entire radio emission may be regarded as a single, large-scale halo. In this view, distinguishing between mini-halo and giant halo components may not be physically meaningful. This interpretation is further supported by the nearly constant spectral index of $\alpha=-1.08\pm0.06$.
     \item 
     The radio halo in Abell~1795  exhibits properties typical of giant radio haloes in  merging clusters: a relatively flat radio spectrum ($\alpha=-1.08\pm0.06$) and a radio power consistent with the established mass–power relation. Such characteristics are generally unexpected in the context of the turbulent re-acceleration model and the observed radio bimodality \citep[e.g.,][]{2009A&A...507..661B}. This may indicate that sufficient turbulent energy can occasionally persist to sustain particle re-acceleration even in dynamically relaxed systems. However, such cases are likely rare, as the turbulent re-acceleration model predicts that giant haloes in relaxed clusters should be much less common than in major mergers.

     \item
      The properties of the Abell 1795 radio halo-- particularly its radio brightness profile closely matching the X-ray profile--and the apparent relaxed dynamical status of the cluster, are very tempting regarding a hypothesis in which the observed emission is generated by secondary electrons. Motivated by this, we modeled the observed radio surface brightness profile in the context of secondary models. In this case the challenge is due to the linear scaling of the radio to X-ray point-to-point brightness which requires radially flat cosmic rays and/or magnetic field energy densities. Specifically we find that assuming a magnetic field scaling with gas density, the observed radio profile is well reproduced by the model under the assumption of a constant energy density of cosmic ray protons with radius and $B {\sim} 1{-}5$~$\mu$G. This challenge requires further exploration of the modeling of a secondary origin of the halo, including the possibility where secondary electrons are re-accelerated.     
     \item
     The discovery of a large radio halo in Abell~1795 may indicate that previous studies were biased against detecting such haloes in cool-core clusters hosting bright central AGN. Deeper, high–dynamic-range radio observations are therefore warranted to build a more complete picture of radio-halo bimodality and its connection to cluster dynamics, in light of this unexpected detection in Abell~1795.

\end{itemize}

\section*{Acknowledgements}
We thank the anonymous reviewer for helpful comments. 
RT is grateful for support from the UKRI Future Leaders Fellowship (grant MR/T042842/1). This work was supported by the STFC [grants ST/T000244/1, ST/V002406/1]. CJR acknowledges financial support from the German Science Foundation DFG, via the Collaborative Research Center SFB1491 `Cosmic Interacting Matters – From Source to Signal'. MB acknowledges support from the Deutsche Forschungsgemeinschaft under Germany's Excellence Strategy - EXC 2121 ``Quantum Universe'' -- 390833306. FDG acknowledges support from the ERC Consolidator Grant ULU 101086378. AB acknowledges support from the ERC CoG $\vec{B}$ELOVED, GA n. 101169773. The Dunlap Institute is funded through an endowment established by the David Dunlap family and the University of Toronto. We thank Lawrence Rudnick for the suggestion to create the radio--X-ray residual images.

The authors acknowledge the OSCARS project, which has received funding from the European Commission’s Horizon Europe Research and Innovation programme under grant agreement No.~101129751.

The MeerKAT telescope is operated by the South African Radio Astronomy Observatory, which is a facility of the National Research Foundation, an agency of the Department of Science and Innovation. This research has made use of data obtained from the Chandra Data Archive provided by the Chandra X-ray Center (CXC).

This manuscript is based on data obtained with the International LOFAR Telescope (ILT). LOFAR \citep{vanhaarlem+13} is the Low Frequency Array designed and constructed by ASTRON. It has observing, data processing, and data storage facilities in several countries, which are owned by various parties (each with their own funding sources), and which are collectively operated by the ILT foundation under a joint scientific policy. The ILT resources have benefited from the following recent major funding sources: CNRS-INSU, Observatoire de Paris and Universit\'e d'Orl\'eans, France; BMBF, MIWF-NRW, MPG, Germany; Science Foundation Ireland (SFI), Department of Business, Enterprise and Innovation (DBEI), Ireland; NWO, The Netherlands; The Science and Technology Facilities Council, UK; Ministry of Science and Higher Education, Poland; The Istituto Nazionale di Astrofisica (INAF), Italy. This research made use of the Dutch national e-infrastructure with support of the SURF Cooperative (e-infra 180169) and the LOFAR e-infra group. The J{\"u}lich LOFAR Long Term Archive and the German LOFAR network are both coordinated and operated by the J{\"u}lich Supercomputing Centre (JSC), and computing resources on the supercomputer JUWELS at JSC were provided by the Gauss Centre for Supercomputing e.V. (grant CHTB00) through the John von Neumann Institute for Computing (NIC). This research made use of the University of Hertfordshire high-performance computing facility and the LOFAR-UK computing facility located at the University of Hertfordshire and supported by STFC [ST/P000096/1], and of the Italian LOFAR IT computing infrastructure supported and operated by INAF, and by the Physics Department of Turin university (under an agreement with Consorzio Interuniversitario per la Fisica Spaziale) at the C3S Supercomputing Centre, Italy. This publication is part of the project LOFAR Data Valorization (LDV) [project numbers 2020.031, 2022.033, and 2024.047] of the research programme Computing Time on National Computer Facilities using SPIDER that is (co-)funded by the Dutch Research Council (NWO), hosted by SURF through the call for proposals of Computing Time on National Computer Facilities.

Artificial Intelligence tool ChatGPT 4.5 was utilised for copy editing and linguistic refinement to ensure clarity. The final manuscript was reviewed and approved by the human authors.

\section*{Data Availability}
The raw data used in this work are publicly available from the LOFAR Long Term Archive (\url{https://lta.lofar.eu}) and the MeerKAT Science Data Archive (\url{https://archive.sarao.ac.za/}), projects SCI-20220822-RV-01, SCI-20190418-KA-01, and SCI-20210212-CR-01. The calibrated radio images in FITS format are available from the corresponding author upon reasonable request.




\bibliographystyle{mnras}
\bibliography{example} 




\appendix

\section{Comparison with previously published LOFAR images}
\label{sec:appendixpreviousLOFAR}
\begin{figure*}
\centering
\includegraphics[width=0.49\textwidth]{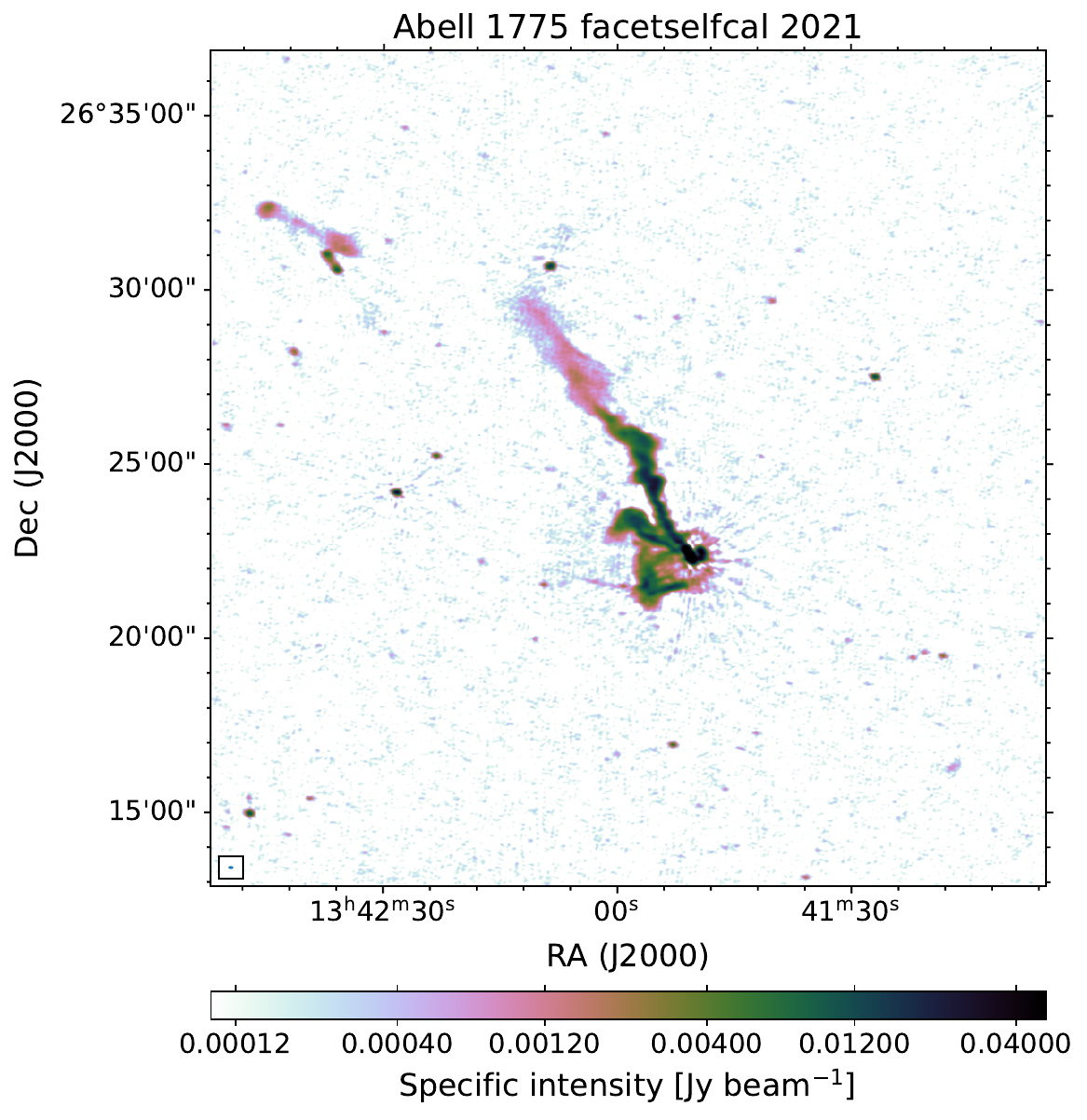}
\includegraphics[width=0.49\textwidth]{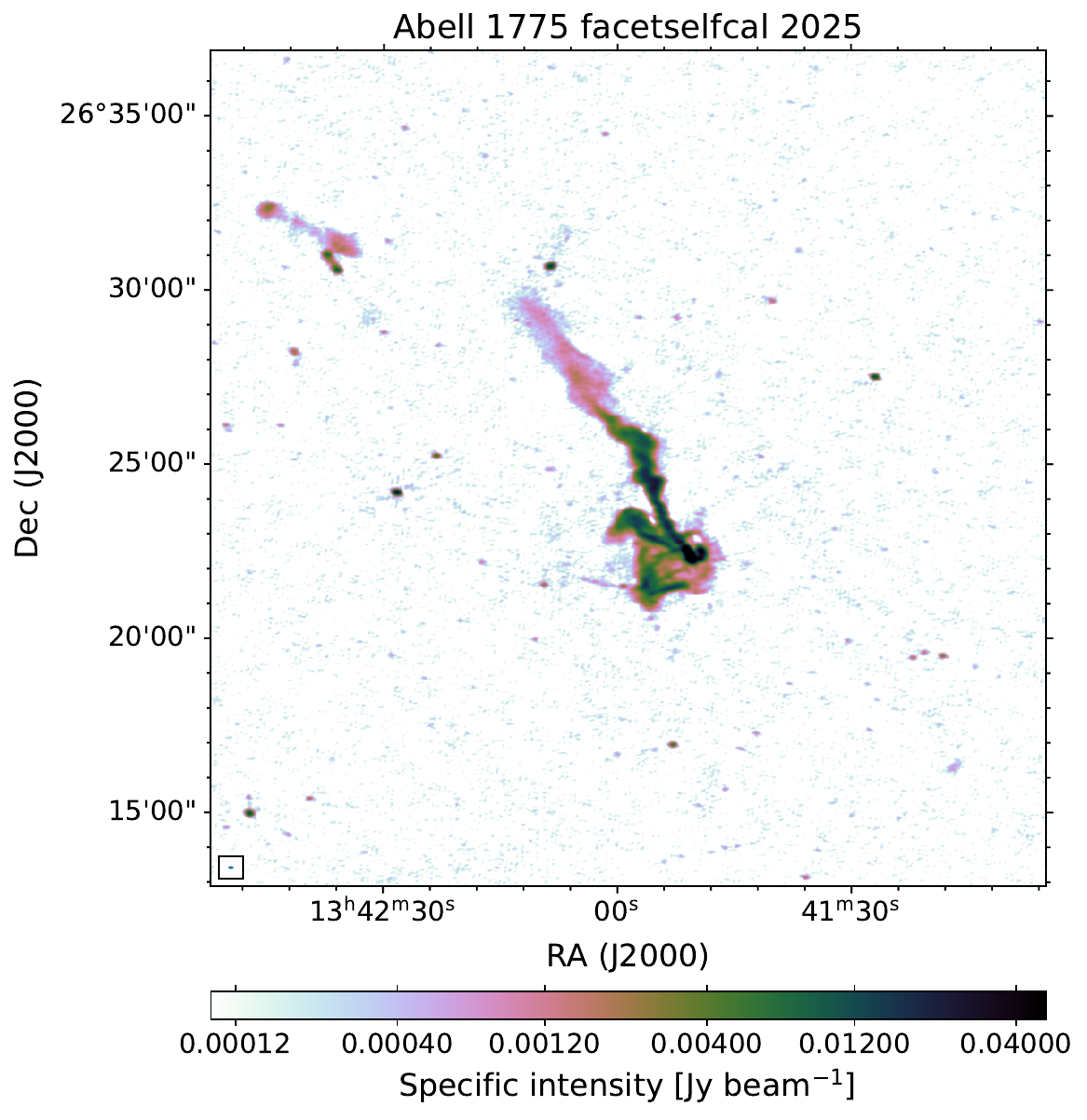}
\includegraphics[width=0.49\textwidth]{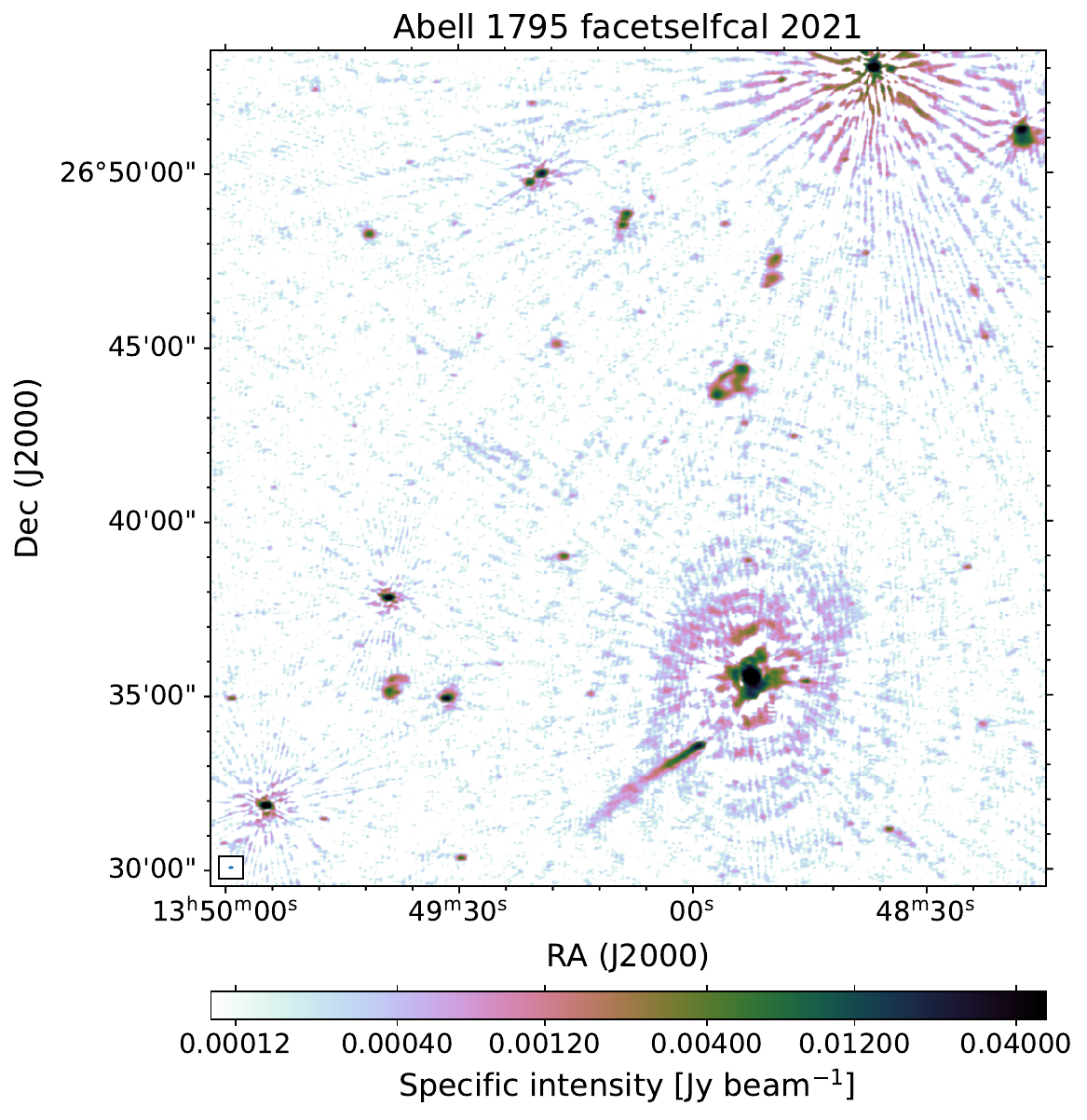}
\includegraphics[width=0.49\textwidth]{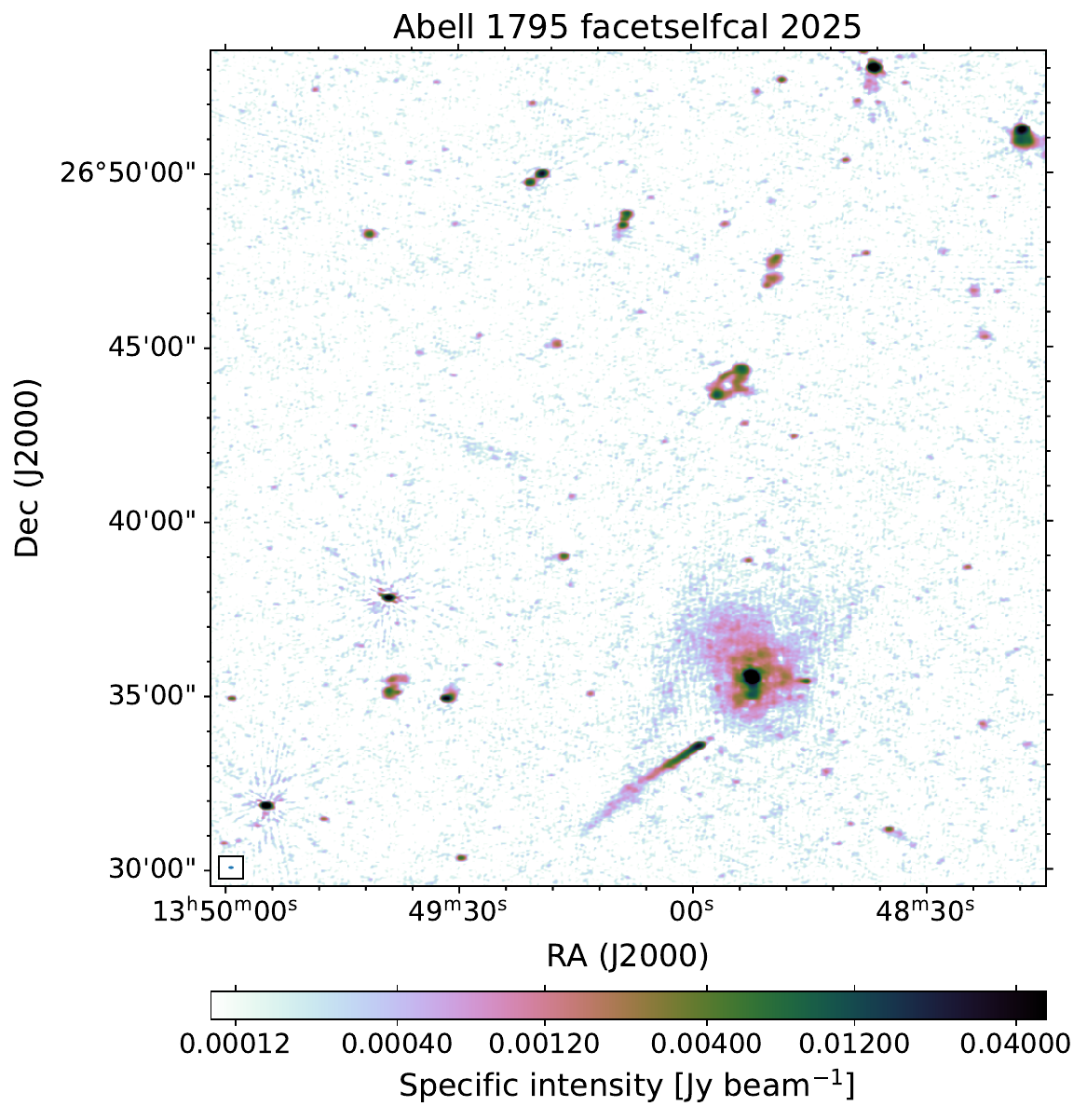}
\caption{Comparison of the image quality in this work using an updated calibration strategy and version of \texttt{facetselfcal} compared to previously published work \citep{2021A&A...649A..37B,2022A&A...660A..78B} which used the strategy presented in \protect\cite{2021A&A...651A.115V}. The top and bottom panels show Abell~1775 and Abell~1795, respectively. The images on the right show the improvement obtained in this work. Images are displayed with identical brightness scaling to aid a direct comparison.}
\label{fig:comp}
\end{figure*}

In Fig.~\ref{fig:comp} we show a comparison between the old published results for Abell~1775 \citep{2022A&A...660A..78B} and Abell~1795 \citep{2021A&A...649A..37B}, using the strategy presented in \cite{2021A&A...651A.115V}, and the improved \texttt{facetselfcal} calibration procedure used in this work.

\section{Spectral index uncertainty maps}
\begin{figure*}
\centering
\includegraphics[width=0.49\textwidth]{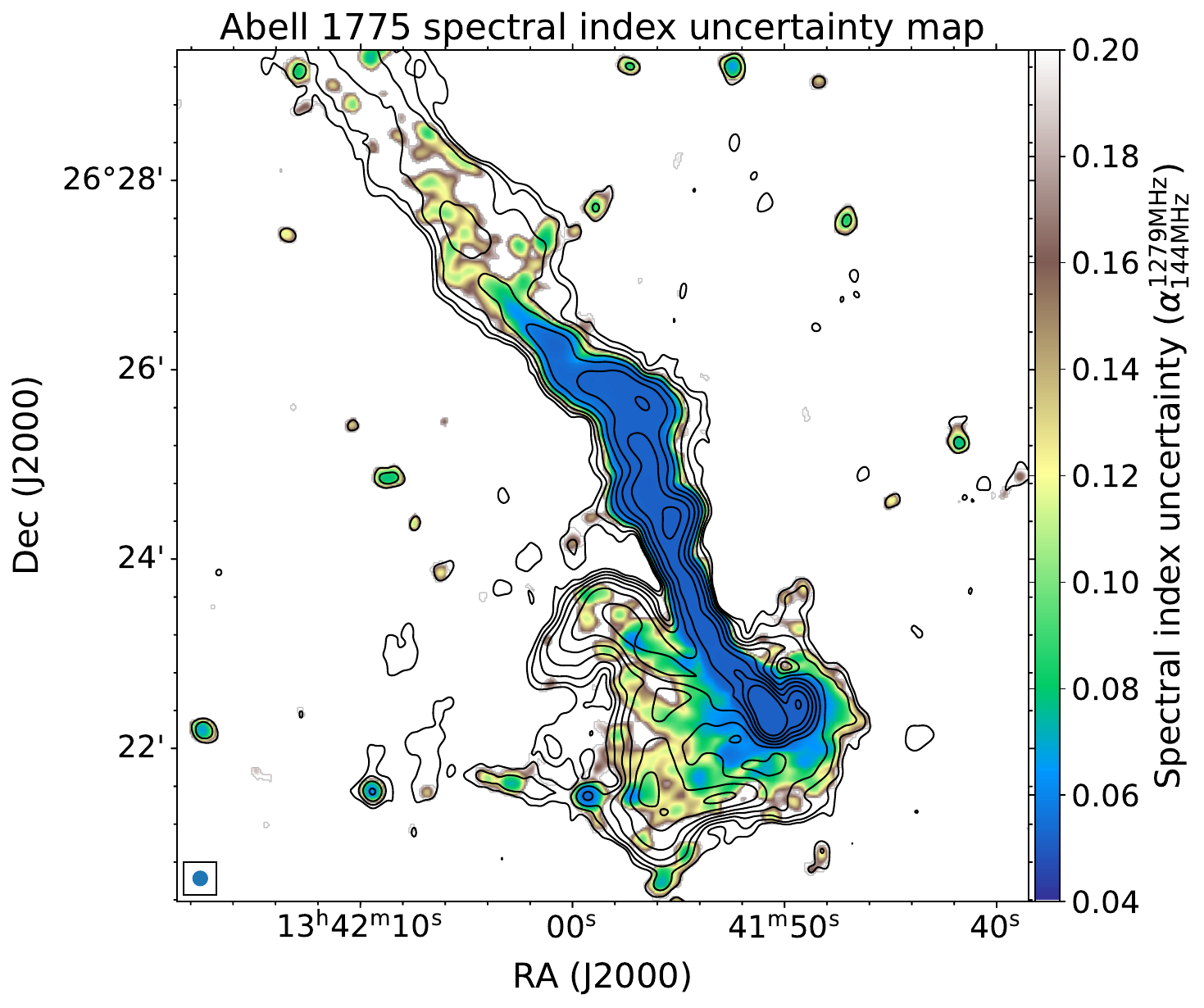}
\includegraphics[width=0.49\textwidth]{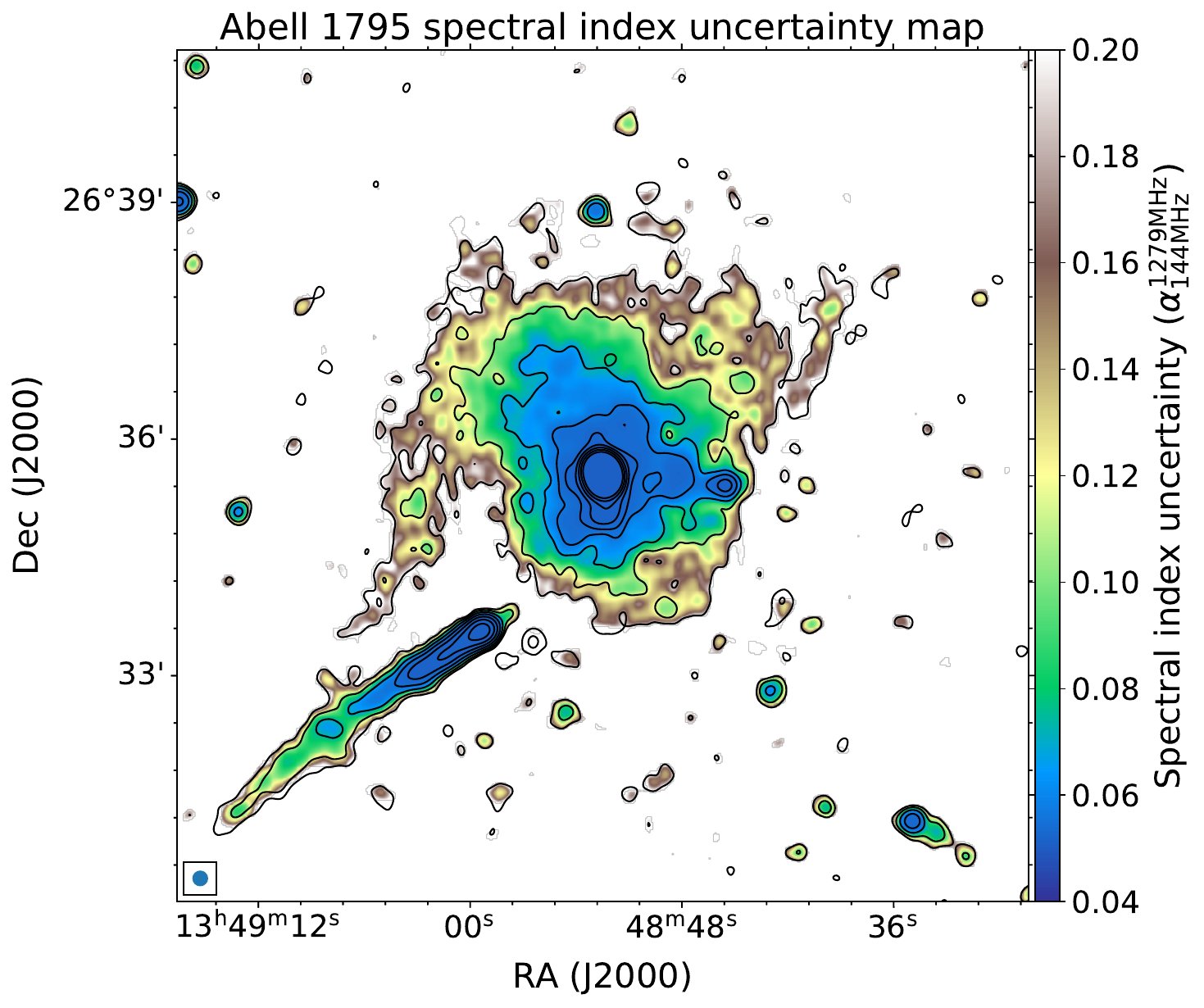}
\caption{Spectral index uncertainty maps corresponding to Figs.~\ref{fig:radioprofileA1775} and~\ref{fig:ChandraMeerKATA1795}, respectively. The spectral index uncertainty includes the r.m.s map noise as well as an absolute flux scale uncertainty of 5\% for MeerKAT and 10\% for LOFAR. The contours are from the 144 MHz LOFAR images and are drawn at the same levels as in  Fig.~\ref{fig:radioprofileA1775} and~\ref{fig:ChandraMeerKATA1795}.}
\label{fig:spixerror}
\end{figure*}

In Fig.~\ref{fig:spixerror} we show the spectral index uncertainty maps corresponding to Fig.~\ref{fig:radioprofileA1775} and~\ref{fig:ChandraMeerKATA1795}.

\section{Low-resolution MeerKAT images}
\label{sec:lowresMeerKAT}
\begin{figure*}
\centering
\includegraphics[width=0.49\textwidth]{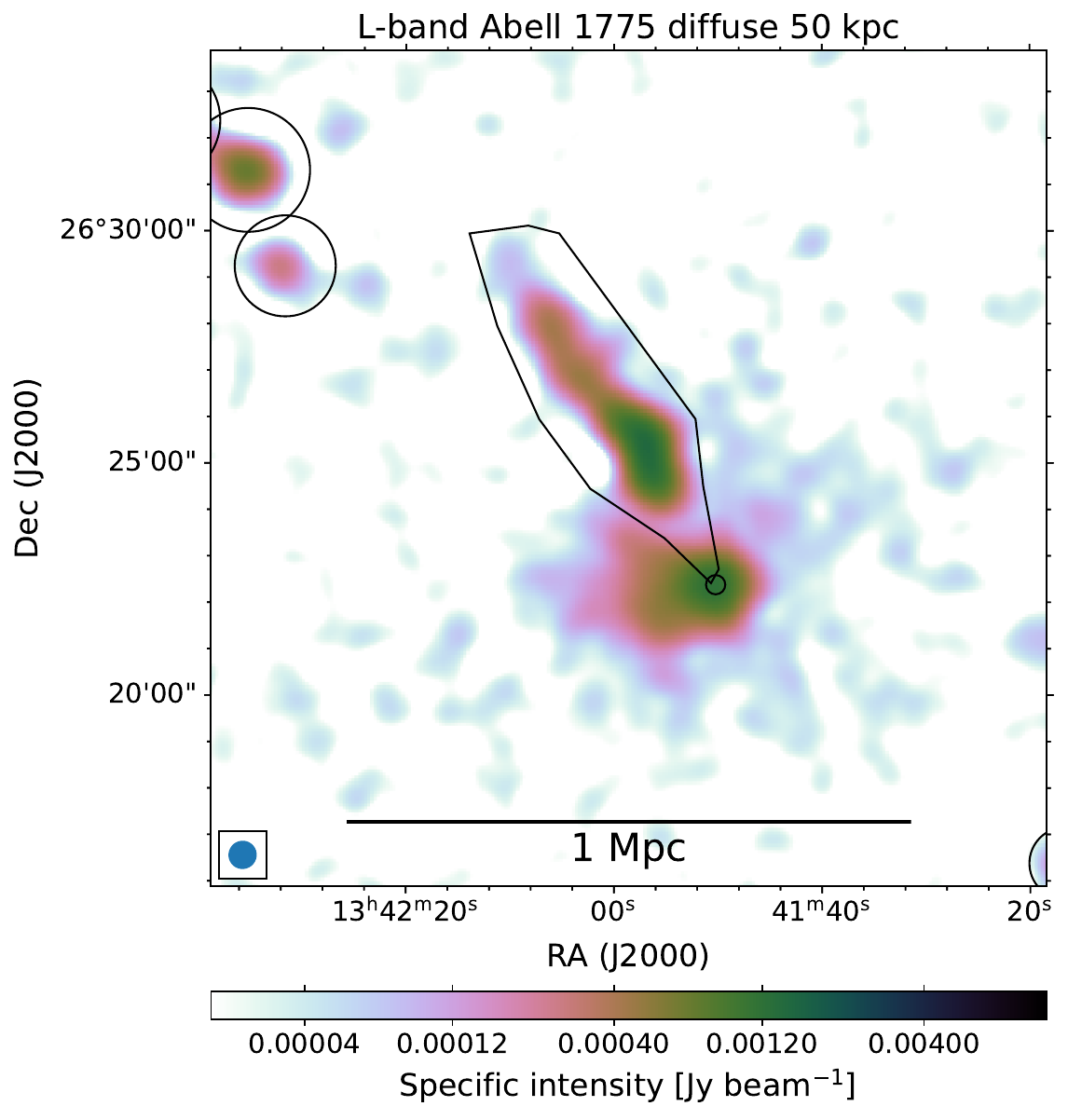}
\includegraphics[width=0.49\textwidth]{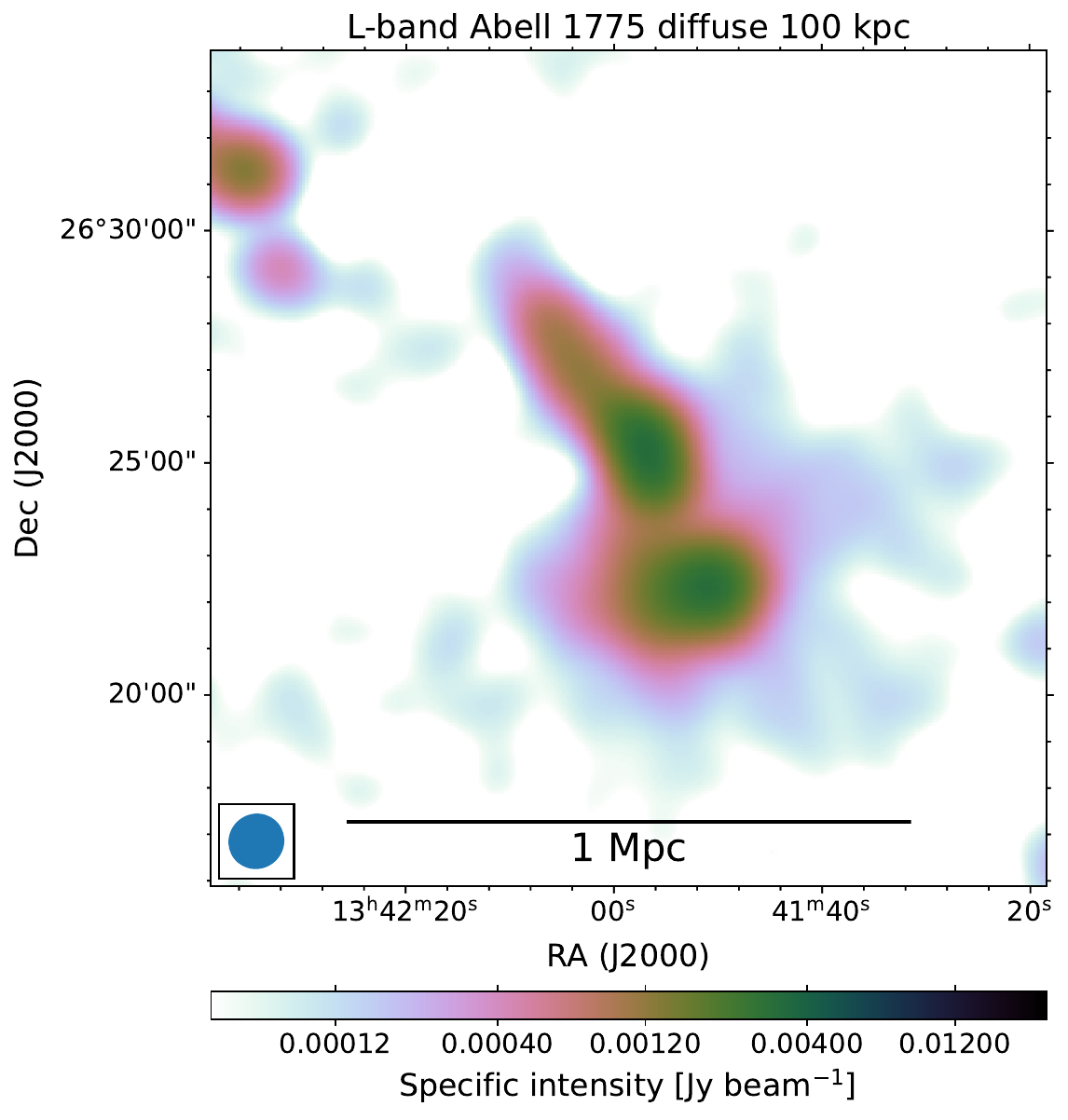}
\includegraphics[width=0.49\textwidth]{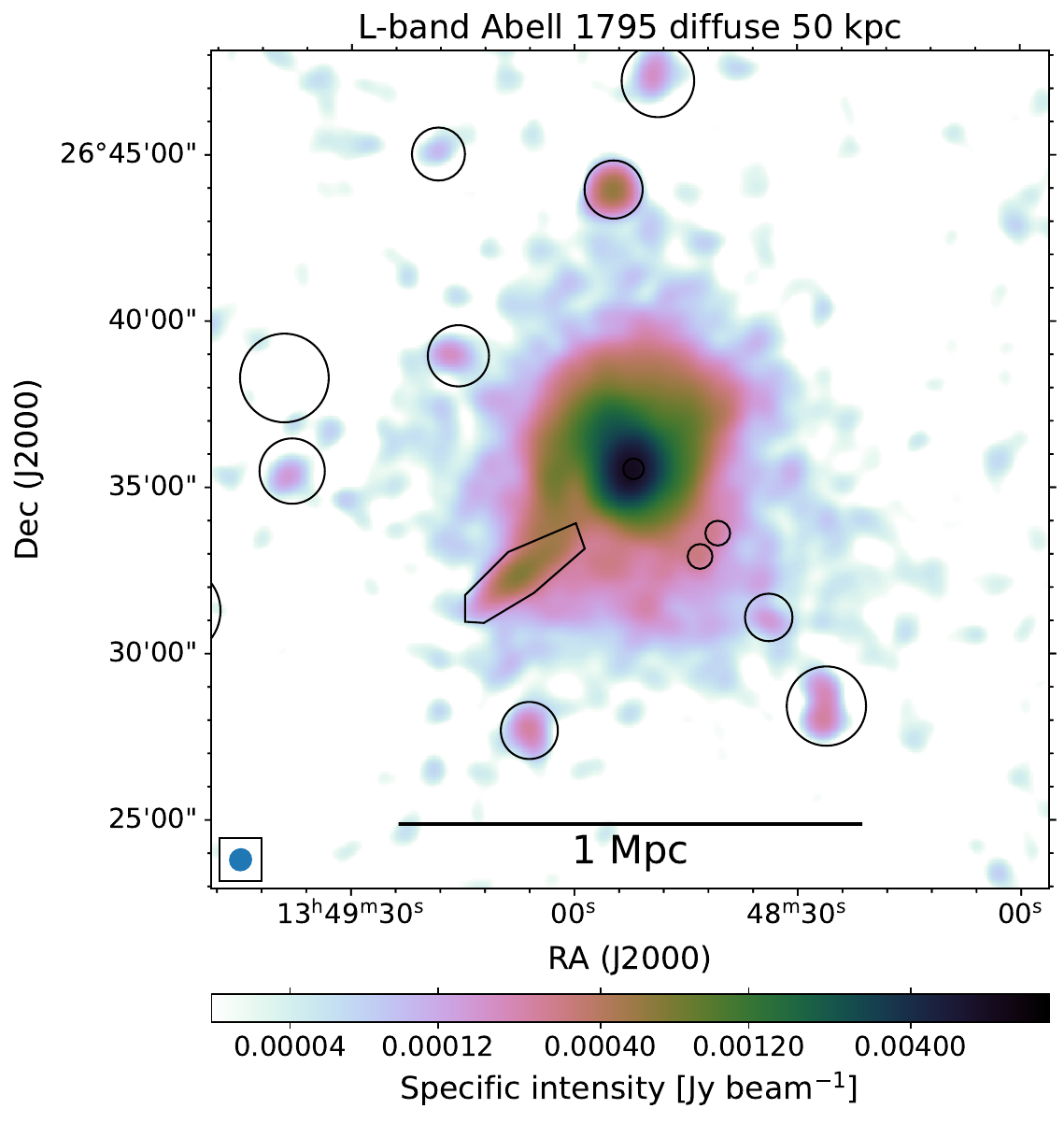}
\includegraphics[width=0.49\textwidth]{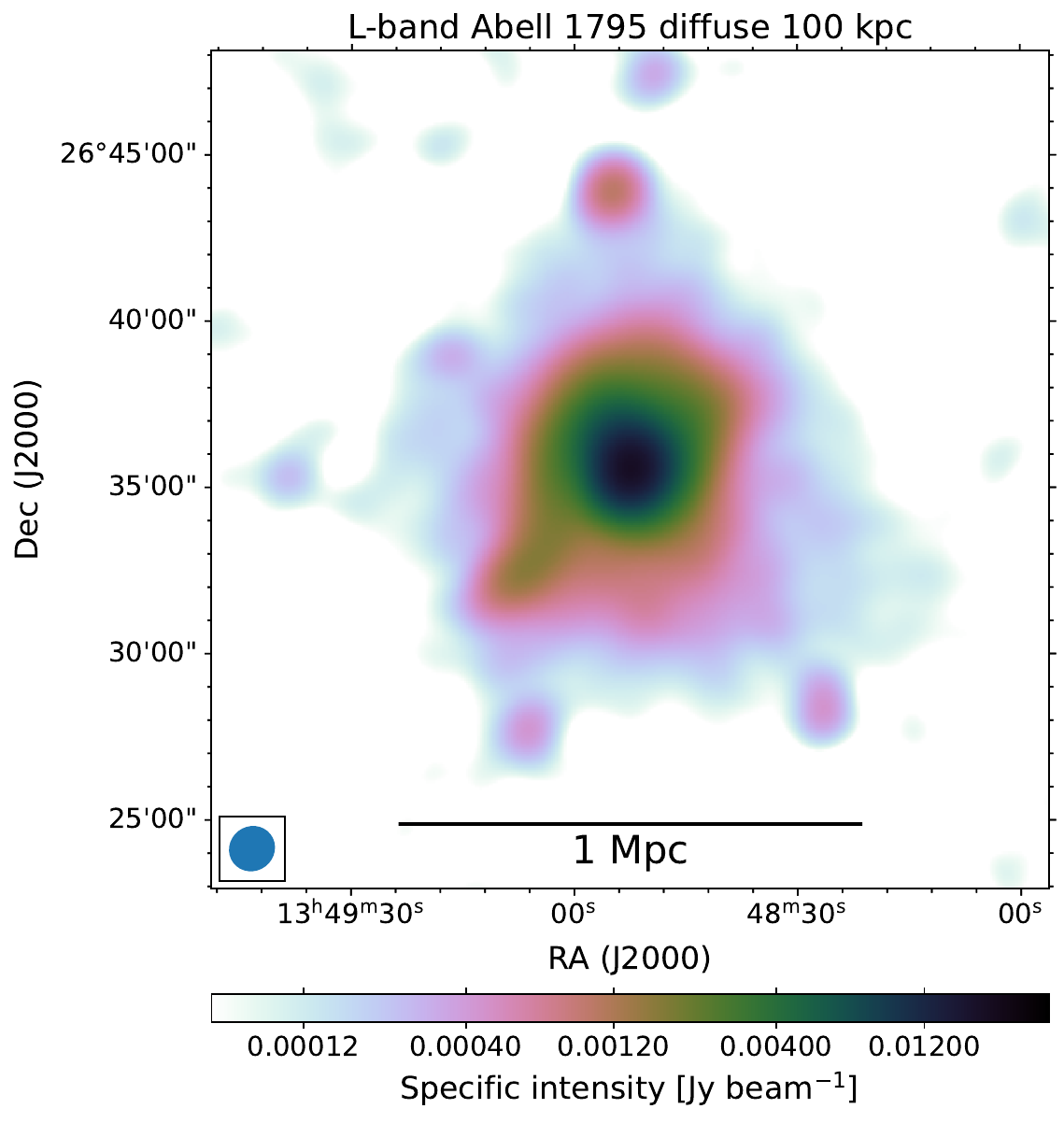}
\caption{MeerKAT 1279~MHz images at 50 and 100~kpc resolution of the diffuse emission in Abell~1775 (top panels) and Abell~1795 (bottom panels). Emission from compact sources was subtracted from the visibilities, see Sect.~\ref{sec:subtraction}. The black regions in the left panels show the areas that were masked to obtain the radial surface brightness profiles displayed in Figs.~\ref{fig:radioprofileA1775} and~\ref{fig:profileA1795}. The noise levels and beam sizes of the radio images are reported in Table~\ref{tab:imageproperties}.}
\label{fig:lowres}
\end{figure*}

In Fig.~\ref{fig:lowres}, we show MeerKAT images of the clusters Abell~1775 (top panels) and Abell~1795  (bottom panels) tapered to a physical resolution of 50 and 100~kpc. The emission from compact sources was subtracted from the visibilities before making these images.

\section{Hadronic radio profiles for different $\eta$-values}
\label{sec:etamodels}

\begin{figure}
\centering
\includegraphics[width=0.48\textwidth]{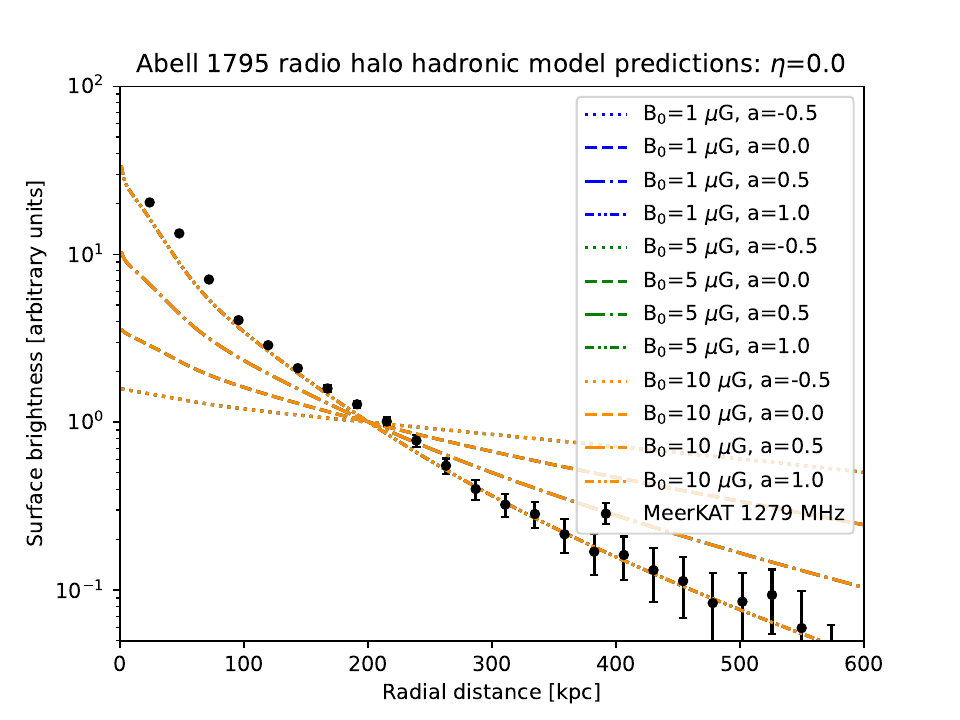}
\includegraphics[width=0.48\textwidth]{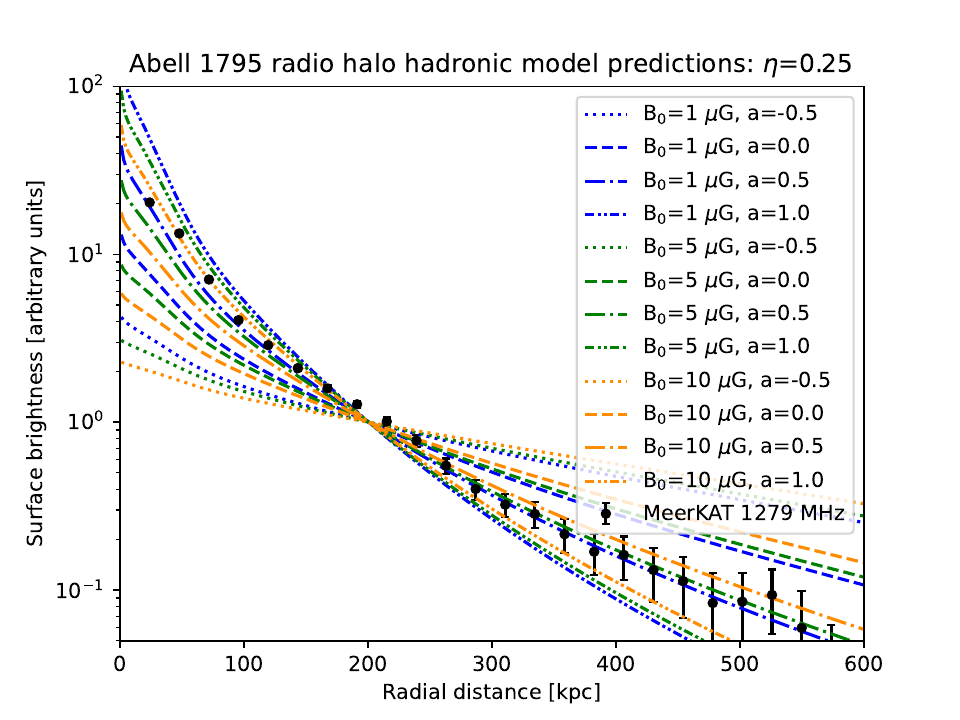}
\includegraphics[width=0.48\textwidth]{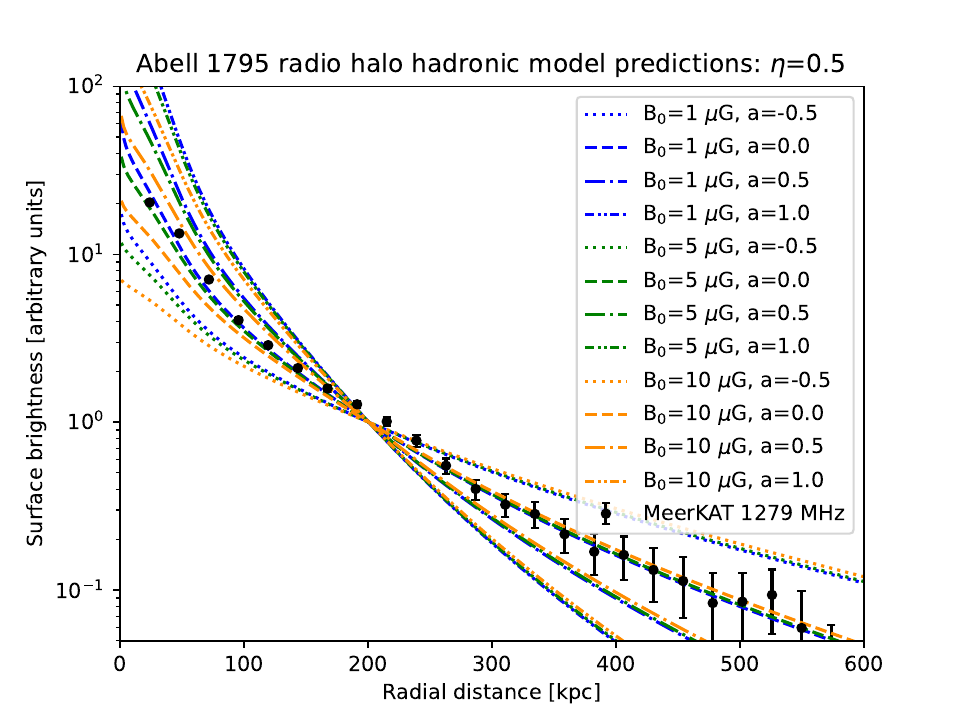}
\caption{Modeled synchrotron surface brightness profiles of the radio halo in Abell~1795, based on hadronic interactions. The profiles are shown for three values of $\eta$, which parameterises the magnetic field scaling as $B(R) \propto n_\mathrm{th}(R)^{\eta}$: $\eta = 0$ (top panel), $\eta = 0.25$ (middle panel), and $\eta = 0.5$ (bottom panel). Black data points indicate measurements from MeerKAT at 1279~MHz. For each panel, profiles are computed for four central magnetic field strengths $B_0$ and three values of the cosmic ray-to-thermal energy density scaling parameter $a$. The profiles are normalised at a radial distance of 200~kpc. For $\eta=0$ (top panel), the profiles for different values $B_0$ overlap. For further details, see Sect.~\ref{sec:hardonic}. }
\label{fig:otheretas}
\end{figure}

Figure~\ref{fig:otheretas} shows modeled radio surface brightness profiles from hadronic interactions for three values of $\eta$, which parameterises the scaling of the cluster magnetic field strength with the thermal gas density (assumed to be $\eta = 0.5$ in Eq.~\ref{eq:Bfield}). The profiles are computed following the procedure described in Sect.\ref{sec:hardonic}.

\bsp	
\label{lastpage}
\end{document}